\documentclass[aps,prb,twocolumn,groupedaddress,showpacs]{revtex4}
\usepackage{graphicx}
\usepackage{amssymb}
\usepackage{amsmath}
\usepackage{bm}
\usepackage{xcolor}

\newcommand{\uu}{\uparrow}
\newcommand{\dd}{\downarrow}

\newcommand{\expect}[1]{\langle{#1}\rangle}
\newcommand{\ket}[1]{|{#1}\rangle}
\newcommand{\bra}[1]{\langle{#1}|}

\newcommand{\abs}[1]{\left|{#1}\right|}

\newcommand{\ttt}{\mathfrak{t}}
\newcommand{\rrr}{\mathfrak{r}}

\newcommand{\Heff}{H_\textrm{eff}} 

\newcommand{\llog}{\text{ln}\,}

\begin{document}

\title{Localization, delocalization, and topological phase transitions
  in the one-dimensional split-step quantum walk}

\author{Tibor Rakovszky} \affiliation{Institute of Physics, 
E\"otv\"os University, P\'azm\'any P\'eter s\'et\'any 1/A, 
H-1117 Budapest, Hungary} 
\author{Janos K. Asboth} \affiliation{Institute for Solid State
  Physics and Optics, Wigner Research Centre for Physics, 
Hungarian Academy of
  Sciences, H-1525 Budapest P.O. Box 49, Hungary}
\date{Spring 2015}
\begin{abstract}
Quantum walks are promising for information processing tasks because
on regular graphs they spread quadratically faster than random
walks. Static disorder, however, can turn the tables: unlike random
walks, quantum walks can suffer Anderson localization, with their
wavefunction staying within a finite region even in the infinite time
limit, with a probability exponentially close to one. It is thus
important to understand when a quantum walk will be Anderson localized
and when we can expect it to spread to infinity even in the presence
of disorder.  In this work we analyze the response of a 
one-dimensional quantum walk -- the split-step walk -- to different
forms of static disorder.  We find that introducing static,
symmetry-preserving disorder in the parameters of the walk leads to
Anderson localization. In the completely disordered limit, however, a
delocalization transition occurs, and the walk spreads subdiffusively
to infinity.  Using an efficient numerical algorithm, we calculate the
bulk topological invariants of the disordered walk, and find that the
disorder-induced Anderson localization and delocalization transitions
are governed by the topological phases of the quantum walk.
\end{abstract}


\maketitle

\section{Introduction}

Discrete-time Quantum Walks\cite{kempe_2003} (or, simply, quantum
walks) are quantum mechanical generalizations of the random
walk. Their hallmark property is that on a regular lattice they spread
quadratically faster than random walks, i.e., ballistically rather
than diffusively. This makes them valuable in quantum search
algorithms\cite{Shenvi03}, or even for general quantum computing
\cite{dtqw_universal}. Experiments on quantum walks range from
realizations on trapped
ions\cite{travaglione_02,roos_ions,schmitz_ion}, cold atoms in optical
lattices\cite{meschede_science,alberti_electric_experiment,
  alberti_nonclassicality}, to light on an optical
table\cite{gabris_prl,schreiber_science,peruzzo_science_2010,
  white_photon_prl,sciarrino_twoparticle,zhao_15}, but there are many
other experimental proposals\cite{rydberg_walk, kalman_09}.

The dynamics of a quantum walk is given by iterations of a unitary
timestep operator, which can always be written in the form
$U=e^{-i\Heff}$, with $\Heff$ a hermitian operator. In this sense, a
quantum walk is a stroboscopic simulator of an effective Hamiltonian
$\Heff$. This is a powerful theoretical concept that allows much of
the physical intuition about lattice systems to be applied to quantum
walks. As an example, consider quantum walks on regular lattices: the
maximum of the group velocity of the effective Hamiltonian translates
directly to the velocity of ballistic expansion of the walk.


In the presence of static (time-independent) disorder, quantum walks
can lose their advantage over random walks in terms of the speed of
spreading: they can undergo Anderson localization, whereby the
mean-squared distance of the walker from the origin stays bounded even
in the infinite-time limit. Besides
theoretical\cite{joye_10,ahlbrecht_11} and numerical
studies\cite{sciarrino_localization_14}, this effect has also been
observed experimentally\cite{gabris_anderson}.

There are special cases where quantum walks can evade Anderson
localization and spread indefinitely even in the presence of static
disorder. Already the simplest one-dimensional quantum walk with angle
disorder presents such a case: rather than being completely localized,
it spreads subdiffusively\cite{obuse_delocalization}. This feature was
explained in Ref.~\onlinecite{obuse_delocalization} by mapping the
effective Hamiltonian of the quantum walk to chiral symmetric quantum
wires (also see Ref.~\onlinecite{zhao2015disordered}). With an eye
towards potential applications of quantum walks, it is important to
understand under what conditions we should expect Anderson
localization, and when delocalized behaviour, of disordered quantum
walks.



One of the key concepts that can help us understand when to expect
Anderson localization in a quantum walk\cite{edge2015localization}, is
that of topological phases. As noted in
Ref.~\onlinecite{kitagawa_exploring}, the effective Hamiltonian
$\Heff$ of a quantum walk on a regular lattice can be engineered to be
that of a topological band insulator\cite{rmp_kane}. If that happens,
bulk--boundary correspondence\cite{rmp_kane} predicts that the quantum
walk will host topologically protected edge states, whose number is
given by a topological invariant of the bulk.  These states can have a
drastic influence on the time evolution of the walker if they have a
large overlap with the initial state.

Quantum walks have been shown to have a broader range of topological
phases than that of their effective Hamiltonian\cite{asboth_prb}:
their topological invariants depend on details of how the timestep is
performed. These invariants can be expressed as winding numbers of the
bulk timestep operator over a part of the
timestep\cite{asboth_2013,asboth_2014,obuse_asboth2015}, or using a
generalization of the scattering theory of topological phases
\cite{fulga_scattering_2012,scattering_walk2014}.  In this respect
quantum walks are representative of the extreme limit of periodically
driven systems\cite{asboth2015edge,akhmerov_majorana_driven,
  rudner_driven,dora_review}, as is the (closely related) quantum
kicked rotator\cite{edge_hall_11}. In the case of the one-dimensional
split-step quantum walk, topologically protected edge states
not predicted by the invariants of the effective Hamiltonian have even
been observed experimentally\cite{kitagawa_observation}; the
corresponding topological invariants have only recently been
identified\cite{obuse_asboth2015}.



In this paper we explore the relation between Anderson localization
and topological phases for one-dimensional split-step quantum
walks. This is a broad family of quantum walks with chiral and
particle-hole symmetries, which contains as a special case the simple
quantum walk of Ref.~\onlinecite{obuse_delocalization}. We use a
cloning procedure to derive the real-space scattering matrix of the
walk, which allows us to give simple and efficient formulas for the
topological invariants as well as for the localization lengths. We
find that uniform disorder in the rotation angles, which does not
break the symmetries, leads to Anderson localization in the generic
case. At maximal angle disorder, however, there is a disorder-induced
delocalization transition and the walk spreads (sub-)diffusively. We
also obtain a simple interpretation of the delocalized behaviour of
the simple quantum walk, by mapping it to a split-step walk at a
boundary between topological phases.  Finally, we explore the effects
of symmetry breaking disorder using phase disorder, i.e.,
position-dependent but time-independent phase factor applied to the
wave function after every timestep. We find that when this disorder
breaks the symmetries of the system it invariably leads to Anderson
localization. 

This paper is structured as follows. In
Sec.~\ref{sec:splitstep_definitions} we remind the reader of the
definition of the split-step quantum walk\cite{kitagawa_exploring}, of
timeframes, and of the symmetries of this quantum walk. In
Sec.~\ref{sec:lyapunov} we derive our main result: the Lyapunov
exponents of the split-step quantum walks, obtained through the
real-space scattering matrix using a cloning procedure.  In
Sec.~\ref{sec:uniform} we apply this tool to treat uniform disorder in
the rotation angles, find Anderson localization or delocalization
depending on the parameters, and give a new perspective on the
delocalization of the simple quantum walk\cite{obuse_delocalization}.
In Sec.~\ref{sec:phase_disorder} we discuss two types of phase
disorder: one which breaks both symmetries of the quantum walk and
thus leads to Anderson localization, and one which breaks only
particle-hole symmetry, and leads to the same type of behaviour as
disorder in the rotation angles. In Sec.~\ref{sec:conclusions} we draw
some conclusions. We also include pedagogically important examples and
calculations in Appendices. In App.~\ref{sec:app_binary} we discuss
the pedagogical case of split-step quantum walks with binary
disorder. In App.~\ref{sec:critical_exponent} we calculate the
critical exponent of the localization-delocalization transition.  In
App.~\ref{sec:realspace_winding} we calculate the topological
invariants of the split-step walk using the noncommutative
generalization of the winding number\cite{mondragon2013topological}.

\section{The Split-step quantum walk}
\label{sec:splitstep_definitions}

In this work we consider the split-step quantum walk on a
one-dimensional chain. The state of the walker at each integer time
$t\in\mathbb{N}$ is represented by a wavefunction,
\begin{align} \ket{\Psi(t)} &= \sum_{x =
    -\infty}^\infty \sum_{s = \uu,\dd} \psi(x,s,t) \ket{x,s}.
\end{align}
Here $x\in\mathbb{Z}$ is the coordinate and $s\in\{\uu,\dd\}$ is the
internal degree of freedom, which we refer to as spin (in the quantum
walk literature, the term ``coin'' is often used).

The quantum walk consists of a sequence of three types of operations.
Spin-dependent shift operations displace the walker, but
do not mix the two spin components,
\begin{align}
  S_\uu &= \sum_{x} \big(\ket{x+1,\uu}\bra{x,\uu} +
  \ket{x,\dd}\bra{x,\dd} \big); \\
  S_\dd &= \sum_{x} \big(\ket{x,\uu}\bra{x,\uu} +
  \ket{x-1,\dd}\bra{x,\dd} \big).
\end{align}
Spin rotation operators rotate the spin about the $y$ axis through a
position-dependent angle,
\begin{align}
  R(\theta) &= \sum_{x} \sum_{s=\uu,\dd} e^{-i\theta(x) \sigma_y} \ket{x,s}\bra{x,s}.
\end{align}
Finally, phase operators multiply the wavefunction by a position- and
spin-dependent phase factor,
\begin{align}
  P(\phi) &= \sum_{x} \sum_{s=\uu,\dd} e^{-i\phi(x,s)} \ket{x,s}\bra{x,s}.
\end{align}

\begin{figure}[h!]
\centering
\includegraphics[width=0.8\columnwidth]{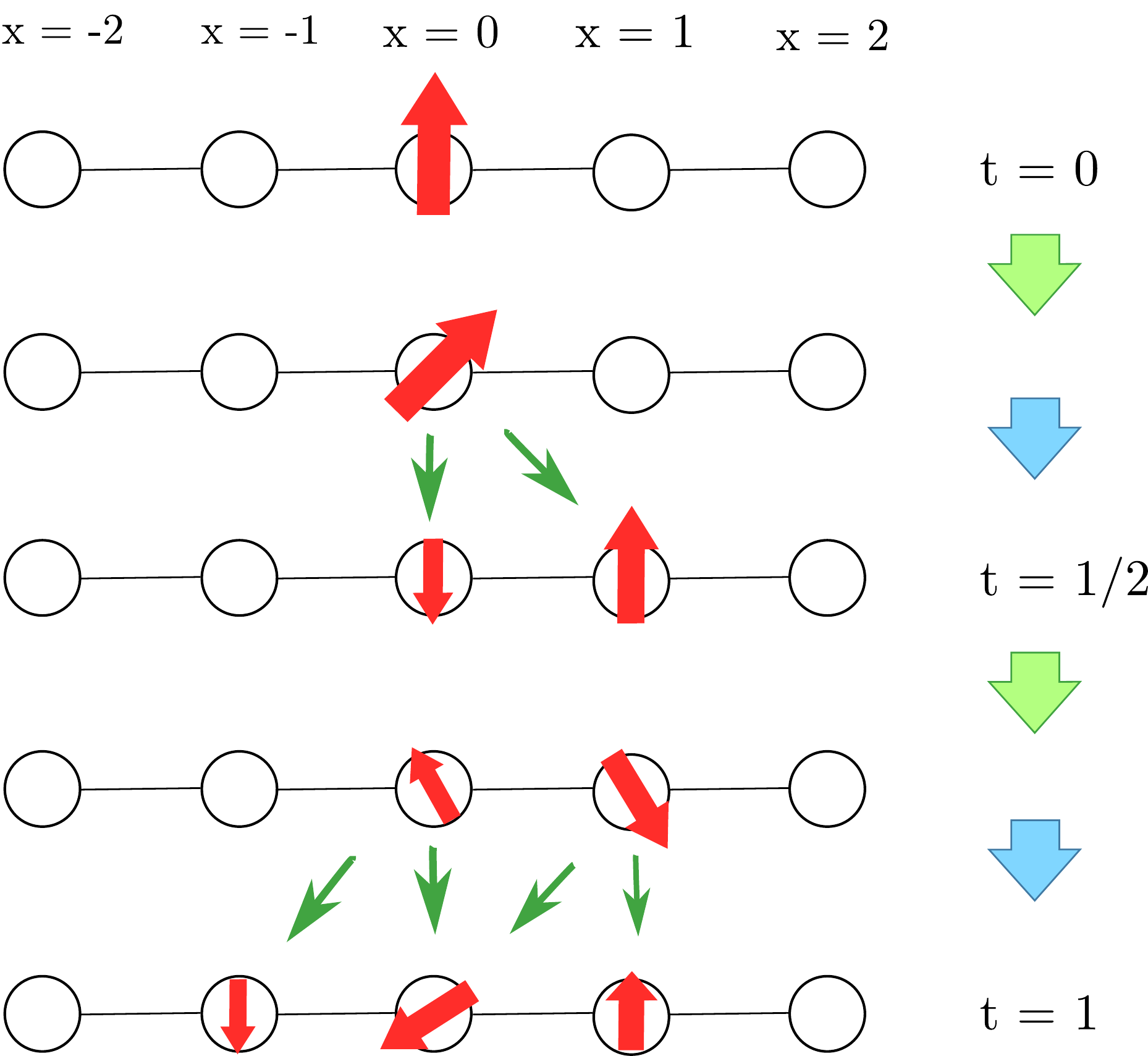}
\caption{(Color online) A full timestep of the split-step walk,
  consisting of 4 operations: (1) rotation of the spin about the $y$
  axis through angle $2\theta_1$, (2) displacement of the $s=+1$
  component of the wavefunction to the right, (3) second spin rotation
  through angle $2\theta_2$, (4) displacement of the $s=-1$ component
  of the wavefunction to the left.}
\label{fig:splitstep}
\end{figure}

The quantum walk is defined by a short sequence of operations which is
then periodically repeated. The effect of this sequence is represented
by the unitary timestep operator, which for the split-step
walk reads
\begin{align}
\label{eq:U_splitstep_def}
U(\theta_1,\theta_2) &= S_\dd R(\theta_2) S_\uu R(\theta_1).
\end{align}
The time evolution is given by 
\begin{align}
\ket{\Psi(t+1)}&= U(\theta_1,\theta_2) \ket{\Psi(t)}, 
\end{align}
as represented in Fig.~\ref{fig:splitstep}.  

The simple quantum walk, as,  e.g., in
Ref.~\onlinecite{obuse_delocalization}, has a single rotation 
operation per period. Its timestep reads 
\begin{align}
\label{eq:U_simple_def}
U_s(\theta) &= S_\dd S_\uu R(\theta). 
\end{align}
This can be seen as a special case of the split-step quantum walk, 
with $\theta_2=0$. 

We will also consider the phase disordered quantum walk. The timestep
operator of this walk reads
\begin{align}
\label{eq:U_p_def}
U_p(\phi,\theta_1,\theta_2) &= P(\phi) S_\dd R(\theta_2) S_\uu R(\theta_1).
\end{align}

\subsection{Timeframes, timestep operators, effective Hamiltonians}

There is a considerable freedom in specifying a quantum walk, i.e., a
periodic sequence of operations, corresponding to shifting the
starting time of the period, which we refer to as changing the
\emph{timeframe} \cite{asboth_2013}. As an example, consider the
operators
\begin{align}
\label{eq:Up_splitstep_def}
U(\theta_1,\theta_2)' &= R(\theta_1/2) S_\dd R(\theta_2) S_\uu R(\theta_1/2);\\
\label{eq:Upp_splitstep_def}
U(\theta_1,\theta_2)'' &= R(\theta_2/2) S_\uu R(\theta_1) S_\dd R(\theta_2/2).
\end{align}
These both correspond to the split-step quantum walk as defined by
Eq.~\eqref{eq:Upp_splitstep_def}, only in different timeframes. 
Timestep operators describing the same quantum walk in different
timeframes are related to each other by a unitary transformation.

The effective Hamiltonian $\Heff$ of a quantum walk is defined as the
logarithm of the unitary timestep operator, 
\begin{align} 
\Heff &= i \llog U.
\end{align} 
The branch cut in the logarithm is taken along the negative real axis,
and so all eigenvalues of $\Heff$, the \emph{quasienergies}
$E$, are between $-\pi$ and $\pi$. Since the unitary
timestep operator $U$ depends on the choice of timeframe (initial time
of the period), the same quantum walk has many, unitary equivalent
effective Hamiltonians associated with it.

\subsection{Symmetries and topological phases}
The split-step walk
has both particle-hole symmetry represented by complex conjugation
$K$, and chiral symmetry, which places the system in Cartan class BDI
\cite{asboth_2013}. 

To see particle-hole symmetry of the quantum walk, notice that all
matrix elements of the timestep operator $U(\theta_1,\theta_2)$ are
real (in position and $\sigma_z$ basis). Thus,
\begin{align}
K U(\theta_1,\theta_2) K &= U(\theta_1,\theta_2),
\end{align}
and therefore $K \Heff K = -\Heff$, which is the defining relation of
particle-hole symmetry. This holds for the split-step quantum walk in
the two timeframes defined by Eqs.~\eqref{eq:Up_splitstep_def} and
\eqref{eq:Upp_splitstep_def} as well. We remark that for a
periodically driven particle-hole symmetric Hamiltonian, the symmetry
is inherited by the effective Hamiltonian in all
timeframes\cite{akhmerov_majorana_driven}.

To see chiral symmetry of a quantum walk explicitly, it is necessary
to go to a chiral symmetric timeframe\cite{asboth_2013}. In the case
of the split-step walk, there are two such timeframes, specified by
Eqs.~\eqref{eq:Up_splitstep_def} and \eqref{eq:Upp_splitstep_def}.  In
these timeframes, we have
\begin{align}
\sigma_x U(\theta_1,\theta_2)' \sigma_x &= {U(\theta_1,\theta_2)'}^\dagger,
\end{align}
and, consequently, $\sigma_x \Heff' \sigma_x = -\Heff'$, which is the
defining relation of chiral symmetry. However, unlike particle-hole
symmetry, the chiral symmetry requirement is nonlocal in
time\cite{asboth_2013}. The addition of a phase shift
operation to the timestep, as in Eq.~\eqref{eq:U_p_def} breaks both
time-reversal and chiral symmetries.

The presence of chiral symmetry enables us to assign bulk topological
invariants to the quantum walk, similar to those in static
systems. Due to the periodicity of the quasienergy, however, edge
states -- eigenstates of the chiral symmetry operator -- can exist at
either $E=0$ or $E=\pi$ quasienergies. This means that there are two
different topological invariants, $\nu_0$ and $\nu_\pi$, associated
with the bulk. 

\begin{figure}[h!]
\centering
\includegraphics[width=0.80\columnwidth]{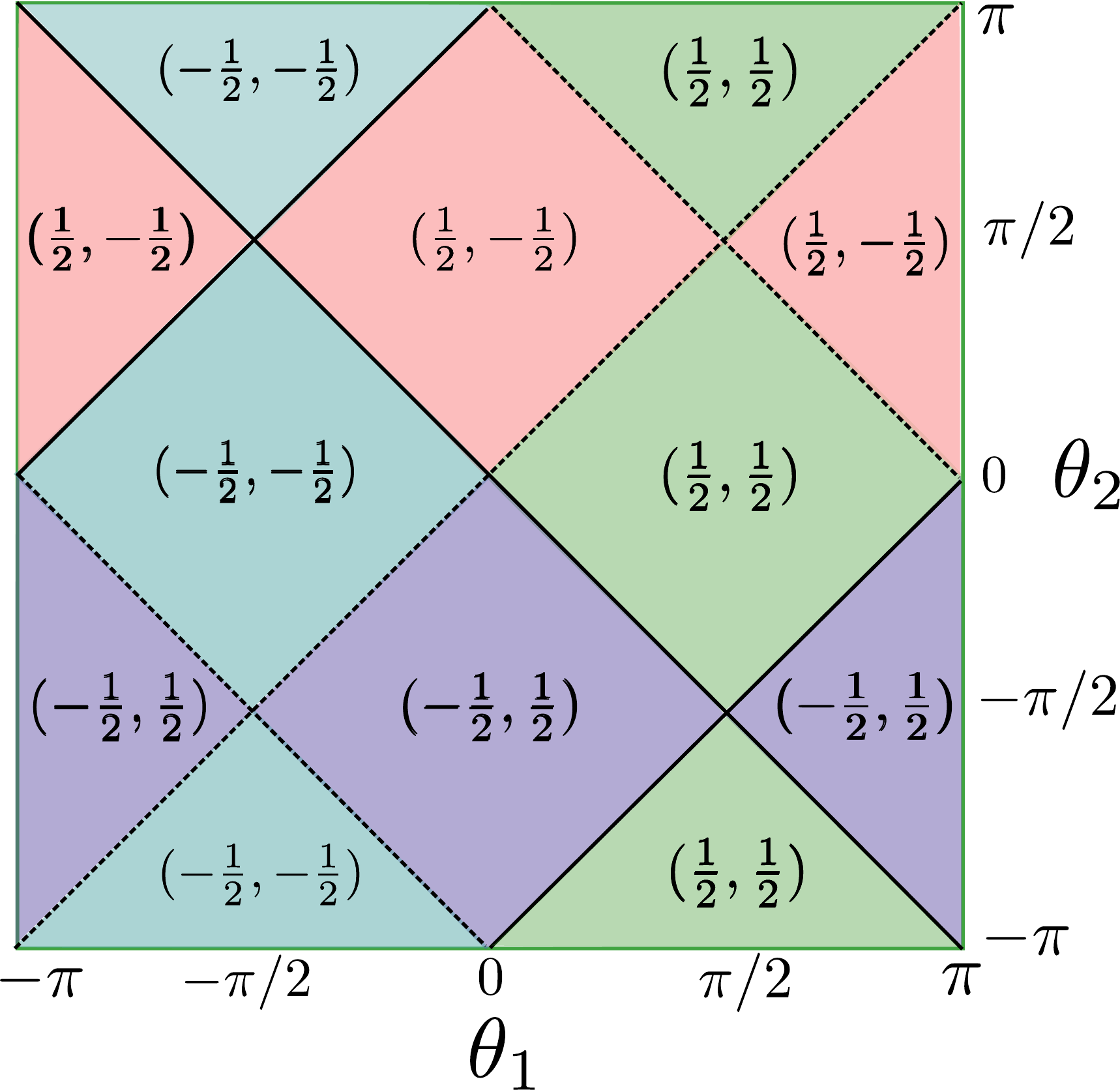}
\caption{(Color online) Phase map of the split-step walk, with gapped phases indexed
  by their pair of bulk topological invariants,
  $(\nu_0,\nu_\pi)$. Along the continuous (dashed) lines, the
  quasienergy gap at $E=0$ ($E=\pi$) closes. }
\label{fig:phasemap}
\end{figure}

In the translation invariant case, when the angles $\theta_1$ and
$\theta_2$ are the same for all coordinates, we can obtain the 
the topological invariants by calculating the winding number in
both chiral timeframes given by \eqref{eq:Up_splitstep_def} and
\eqref{eq:Upp_splitstep_def}. These two winding numbers can be
combined to give the topological invariants $\nu_0$ and $\nu_\pi$
\cite{asboth_2013}, resulting in the topological phase map shown in
Fig.~\ref{fig:phasemap}

\section{Lyapunov exponents and topological invariants by cloning}
\label{sec:lyapunov}

In this paper we will be concerned with the effect of disorder on the
topological invariants and edge states of the split-step walk.  Since
disorder breaks translation invariance, the bulk topological
invariants can no longer be obtained as $k$-space winding
numbers. There are two alternative approaches to the topological
invariants for the disordered case: the one based on the scattering
matrix \cite{fulga_scattering_2012,scattering_walk2014}, and the one
based on a reformulation of the winding number in real space, recently
used for the disordered SSH model\cite{mondragon2013topological}.  In
the following we detail the first approach. In
App.~\ref{sec:realspace_winding} we briefly describe the second
approach, and compare the results obtained via these approaches.

\begin{figure}[h!]
\centering
\includegraphics[width=0.95\columnwidth]{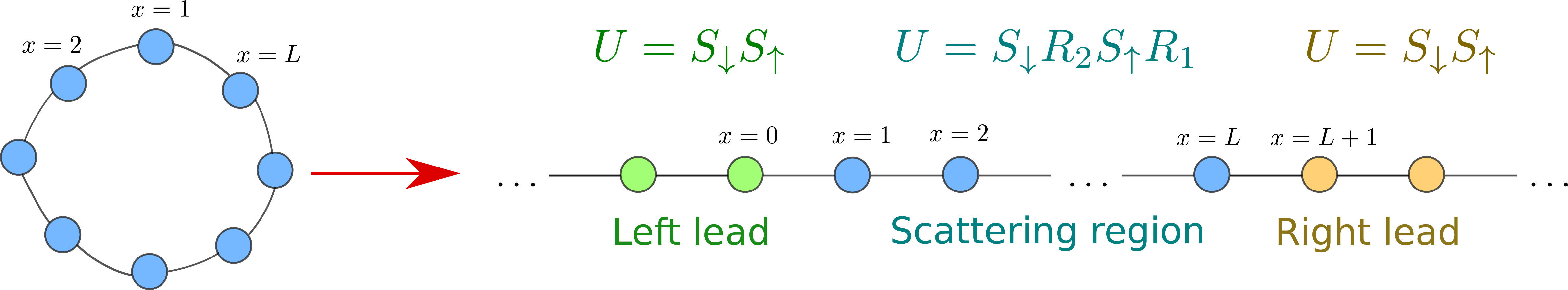}
\caption{(Color online) To implement the scattering matrix for the quantum walk, we
  break up the periodic boundary conditions and attach two leads to
  the two ends of the system. The eigenstates in the leads are plane
  waves with linear dispersion relations.}
\label{fig:leads}
\end{figure}

To define a scattering matrix, we have to use open boundary
conditions, with two translationally invariant leads attached to the
scattering region\cite{scattering_walk2014}. To obtain leads that host
the right number of propagating modes, we omit the rotations in
semi-infinite parts of the system, so the timestep operator there
simply reads $U = S_\dd S_\uu$, as shown in Fig.~\ref{fig:leads}.

The scattering theory of topological invariants gives a simple
way to write down the topological invariants of a bulk: they are
related to the reflection amplitudes of the bulk at the quasienergies
of the edge states. Chiral symmetry ensures that the reflection
amplitudes at these energies are real and the topological invariants
for a bulk are given by\cite{scattering_walk2014}
\begin{equation}
\label{eq:invariants_from_scattering}
	\nu_{E} = \frac{1}{2} r(E). \qquad
        (E=0,\pi)
\end{equation}

The scattering matrix of a quantum chain of $L$ ``slices'' is usually
obtained from the transfer matrix, which is the product of the
$L$ transfer matrices representing the effect of each slice. The
product structure of the transfer matrix makes it straightforward to
obtain the statistical properties of a system with uncorrelated
disorder, if the transfer matrix of each slice is obtained from
variables that are local to the site.

There is a problem with applying the transfer matrix method directly
to a split-step walk. During one timestep the walker makes excursions
to nearest-neighbor sites and thus is not only affected by variables
that are local to one site. 

Our approach to dealing with this problem involves an extension of the
Hilbert space, by introducing $N$ copies, or ``clones'' of the
original split-step walk consisting of $N$ substeps. We explain the
details below for the $N=2$ case, the generalization is
straightforward.


\subsection{Cloning}

\begin{figure}[h!]
 \centering
 \includegraphics[width=0.9\columnwidth]{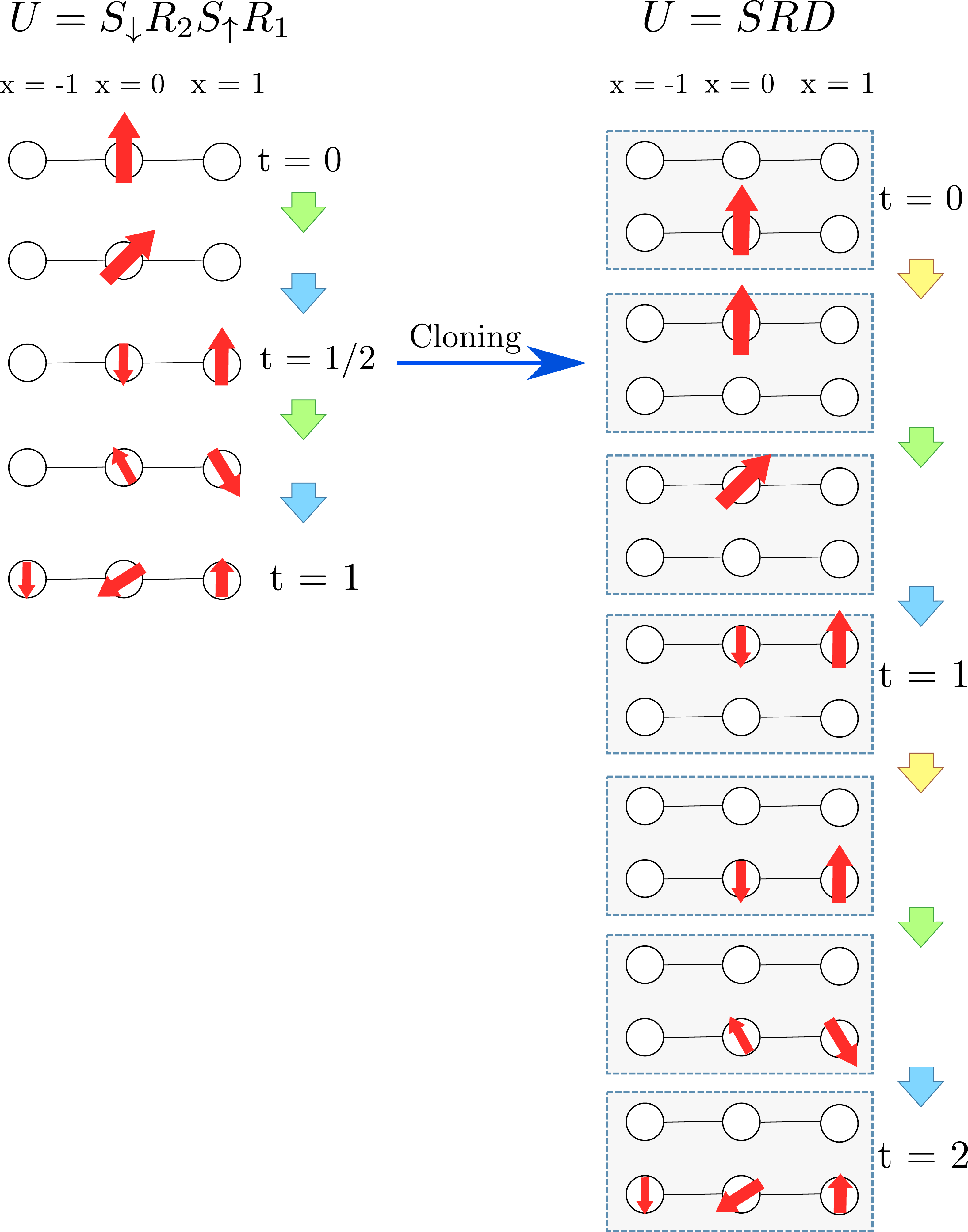}
 \caption{(Color online) Cloning of the split-step walk. Starting from
   the second clone, the state of the walker after $2t$ timesteps of
   the cloned walk is equivalent to the state of the original walker
   after $t$ timesteps.}
 \label{fig:cloning}
 \end{figure}

The central idea of cloning, as shown in Fig.~\ref{fig:cloning}, is
that we double the number of internal degrees of freedom at each site,
thus creating two clones, and we break up the time evolution between
these clones. The first half of the step takes place on the first
clone and the second part on the second clone. At the beginning of
each step, the walker is shifted from one clone to the other by an
operator $D$. In formulas,
\begin{subequations}
\label{eq:cloned_timestep}
\begin{align}
& U_\text{cloned} = SRD \\
S &= S_\uu \bigoplus S_\dd \\
R &= R_1 \bigoplus R_2 \\
D &= \sum_x \sum_{\sigma = \uparrow,\downarrow} 
|x,\sigma,1\rangle\langle x,\sigma,2| + 
|x,\sigma,2\rangle\langle x,\sigma,1|
\end{align}
\end{subequations}
where the indices $1$ and $2$ refer to the first and second clone,
respectively. This definition ensures that on each clone,
$(U_\text{cloned})^2$ is the same as the original timestep in some
timeframe depending on which clone the walker starts from. 

In order to have the original time evolution, we need to start the
particle from the second clone. In that case, each step in the cloned
walk is equivalent to half a step in the original split-step walk
defined in Eq.~\eqref{eq:U_splitstep_def},
\begin{equation}
\label{eq:cloned_splitstep}
U_\text{cloned}^2 = U_\text{splitstep}.
\end{equation}

The main advantage of cloning is that the walker cannot return to the
site from which it started in a single step. This enables us to write
down simple formulas for the stationary states. Note that according to
Eq.~\eqref{eq:cloned_splitstep} the eigenstates of the original split
step walk with quasienergy $E$ will appear as states with quasienergy
$E/2$ in the cloned walk.

In general, a state in the cloned system has the form
\begin{multline}
\ket{\psi} = \sum_x\sum_{s = \uparrow,\downarrow} 
\psi(x,s,1)|x,s,1\rangle + \psi(x,s,2) \ket{x,s,2}
\end{multline}
The $\psi(\uparrow,2)$ and $\psi(\downarrow,1)$ components of the
wavefunctions are unaffected by the shift operation $S$, while the
$\psi(\uparrow,1)$ component is shifted one site to the right and
$\psi(\downarrow,2)$ is shifted one site to the left (both these
states are also shifted to the other clone). Thus we can divide the
wavefunction into three parts,
\begin{subequations}
\begin{align}
\psi_x^\uparrow &= \psi(x,\uparrow,1), \\ 
\psi_x^\downarrow &= \psi(x,\downarrow,2), \\
\psi_x^0 &=  (\psi(x,\downarrow,1)\, ,\,\psi(x,\uparrow,2)).
\end{align}
\end{subequations}

We then define the matrix $A$ to be the part of the timestep operator
preceding the shift in the basis where the wavefunction is ordered as
$\psi_x = (\psi_x^\uparrow\, ,\,\psi_x^\downarrow\, ,\, \psi_x^0)$:
\begin{equation}
A \equiv RD =
\begin{pmatrix}
	A^{\uparrow\uparrow} & A^{\uparrow\downarrow} & A^{\uparrow 0} \\
	A^{\downarrow\uparrow} & A^{\downarrow\downarrow} & A^{\downarrow 0} \\
	A^{0\uparrow} & A^{0\downarrow} & A^{00}
\end{pmatrix}
= \begin{pmatrix}
A^{\uu} \\
A^{\dd} \\
A^0
\end{pmatrix},
\end{equation}
where we introduced $A^{mn}$ that maps from sector $n$ of the
wavefunction to sector $m$, with $m,n \in \{\uparrow,\downarrow,0\}$,
and $A^{m} \equiv (A^{m\uu}, A^{m\dd}, A^{m 0})$. Each component
depends on $x$.

The equation for the components of the scattering states can be
written as
\begin{equation}
\label{eq:stationary_condition}
\begin{pmatrix}
A_{x-1}^{\uu}\Psi_{x-1}^{\uu} \\
A_{x+1}^{\dd}\Psi_{x+1}^{\dd} \\
A_{x}^0\Psi_x^0
\end{pmatrix}
= e^{-i\varepsilon}
\begin{pmatrix}
\Psi_x^{\uu} \\
\Psi_x^{\dd} \\
\Psi_x^0
\end{pmatrix},
\end{equation}
where $\varepsilon = E/2$ is the quasienergy measured in the cloned
system. The last component of Eq.~\eqref{eq:stationary_condition} can
be solved for $\psi_x^0$,
\begin{equation}
\label{eq:psi0}
\Psi_x^0 = G_x(\varepsilon)(A_x^{0\uu}\Psi_x^{\uu} + A_x^{0\dd}\Psi_x^{\dd}),
\end{equation}
where we introduced the shorthand
\begin{equation}
G_x(\varepsilon) \equiv (e^{-i\varepsilon}1\!\!1_{2\times 2} - A_x^{00})^{-1}
\end{equation}
for the resolvent of the matrix $A_x^{00}$. This enables us to relate
$\Psi^{\uu}$ and $\Psi^{\dd}$ on adjacent sites.

\subsection{Real-space scattering and transfer matrix}

We arrive at what can be called the real-space scattering matrix from
Eq.~\eqref{eq:stationary_condition}, using Eq.~\eqref{eq:psi0},  
\begin{equation}
\label{eq:realspace_smatrix_def}
\begin{pmatrix}
	\Psi_{x-1}^\downarrow \\
	\Psi_{x+1}^\uparrow
\end{pmatrix} =
\begin{pmatrix}
	\rrr_x & \ttt'_x \\
	\ttt_x & \rrr'_x
\end{pmatrix} \begin{pmatrix}
	\Psi_x^\uparrow \\
	\Psi_x^\downarrow
\end{pmatrix}, 
\end{equation}
with the components of the above matrix defined as 
\begin{subequations}
\label{eq:realspace_scattering_components}
\begin{align}
\rrr_x &= e^{i\varepsilon} \left( A_x^{\downarrow\uparrow} + A_x^{\downarrow 0}
G_x(\varepsilon) A_x^{0^\uparrow} \right);\\
\ttt_x &= e^{i\varepsilon} \left( A_x^{\uparrow\uparrow} + A_x^{\uparrow 0}
G_x(\varepsilon) A_x^{0^\uparrow} \right);\\
\ttt'_x &= e^{i\varepsilon} \left( A_x^{\downarrow\downarrow} + A_x^{\downarrow 0}
G_x(\varepsilon) A_x^{0^\downarrow} \right);\\
\rrr'_x &= e^{i\varepsilon} \left( A_x^{\uparrow\downarrow} + A_x^{\uparrow 0}
G_x(\varepsilon) A_x^{0^\downarrow} \right).
\end{align}
\end{subequations}
We can interpret Eq.~\eqref{eq:realspace_smatrix_def} in the following
way: the components $\Psi_x^\uu$ and $\Psi_x^\dd$ act as incoming
''modes'' that come to the site $x$ during a specific time step from
the left/right respectively, while $\Psi_{x-1}^\dd$ and
$\Psi_{x+1}^\uu$ act as outgoing modes towards the left/right from the
same site. This justifies the name real-space scattering matrix for
the matrix appearing in Eq.~\eqref{eq:realspace_smatrix_def}.

We can now obtain the reflection and transmission amplitudes that
relate out- to ingoing plane waves in the two leads. 
First, we combine the $L$ individual $2\times2$ real-space scattering
matrices of the sites by the usual combination rule of scattering
matrices, and thus get the real-space scattering matrix of the whole
scattering region.  From the components $\rrr_{1,L}$ and $\ttt_{1,L}$
of that matrix, the reflection and transmission amplitudes are given
by the following formulas,
\begin{subequations}
\begin{align}
\label{eq:reflection_from_realspace_smatrix}
r(\varepsilon) &= e^{i\varepsilon} \rrr_{1,L} ;\\
t(\varepsilon) &= e^{i\varepsilon} \ttt_{1,L}.
\end{align}
\end{subequations}

The cloning of the system and the real-space scattering matrices give
us a useful method for calculating scattering amplitudes and thus
topological invariants both numerically and analytically. This method
can be easily generalized to more complicated quantum walk protocols
involving multiple shift and rotation operators, or larger number of
internal states. Later, in Sect.~\ref{sec:phase_disorder} we will
use it for the phase disordered split-step walk as well.

Although for numerical work, the real-space scattering matrix is a
practical tool because of its numerical stability, for analytical
formulas, the real-space transfer matrix $M_x$ is more useful. It is
defined by the relation
\begin{equation}
\begin{pmatrix}
\Psi_{x+1}^{\uu} \\
\Psi_{x}^{\dd}
\end{pmatrix}
= M_x \begin{pmatrix}
\Psi_{x}^{\uu} \\
\Psi_{x-1}^{\dd}
\end{pmatrix}
\end{equation}
For the split-step walk, the real-space transfer matrix at quasienergy
$E$ depends on the local angle parameters $\theta_j(x)$ through their
sines and cosines, abbreviated as
\begin{align}
s_j &= \sin \theta_j(x);& c_j &= \cos \theta_j(x),
\end{align} 
with $j=1,2$. 
The formula for the transfer matrix reads 
\begin{multline}
\label{eq:tmatrix}
M_x(E) =
\frac{1}{c_1 c_2} \cdot \\
\begin{pmatrix}
e^{iE} + s_1 s_2 & - e^{-iE/2} s_1 - e^{iE/2} s_2 \\
- e^{iE/2} s_1 - e^{-iE/2} s_2 & - e^{-iE} + s_1 s_2
\end{pmatrix}.
\end{multline}

\subsection{Lyapunov exponents, topological invariants, and localization 
length of the disordered split-step quantum walk}

To determine the topological invariants of the disordered split-step
walk, we need the reflection amplitudes at quasienergies $0$ and
$\pi$, as per Eq.~\eqref{eq:invariants_from_scattering}. At these
quasienergies, the real-space transfer matrix of a single site,
Eq.~\eqref{eq:tmatrix}, is considerably simplified,
\begin{subequations}
\label{eq:tmatrix_parametrization}
\begin{align}
M_x(0) &= e^{\lambda_x(0)\sigma_x} \\
M_x(\pi) &= V e^{\lambda_x(\pi)\sigma_x}
V^{-1},
\end{align}
\end{subequations}
where $V = V^{-1} = (\sigma_x - \sigma_y)/\sqrt{2}$ is a $2\times2$
unitary matrix.  The parameters $\lambda_x(0)$ and $\lambda_x(\pi)$
are functions of the rotation angles $\theta_{1,2}(x)$ of the site,
\begin{subequations}
\label{eq:Lyapunov_exponents}
\begin{align}
\lambda_x(0) = \frac{1}{2} \llog{
\frac{(1-\sin\theta_1(x))(1 - \sin \theta_2(x))}
{(1+\sin{\theta_1(x)})(1 + \sin{\theta_2(x)})}}; \\
\lambda_x(\pi) = \frac{1}{2} \llog{
\frac{(\sin{\theta_1(x)} - 1)(1 + \sin{\theta_2(x)})}
{(1 + \sin{\theta_1(x)})(\sin{\theta_2(x)} - 1)}}.
\end{align}
\end{subequations}

As seen from the Eqs.~\eqref{eq:tmatrix_parametrization}, at the
quasienergies $E=0,\pi$ the real-space transfer matrices of the single
sites all commute. Thus, the quantities $\lambda_x(0)$ and
$\lambda_x(\pi)$ are additive: A system of $L$ consecutive sites is
characterized by their sum, or equivalently, their average,
\begin{align}
\lambda_E &= \frac{1}{L} \sum_x \lambda_x(E) \qquad (E = 0,\pi). 
\end{align}
We will refer to this average $\lambda_E$ as the Lyapunov exponent.  

\begin{figure}[h!]
\centering
\includegraphics[width=0.37\textwidth]{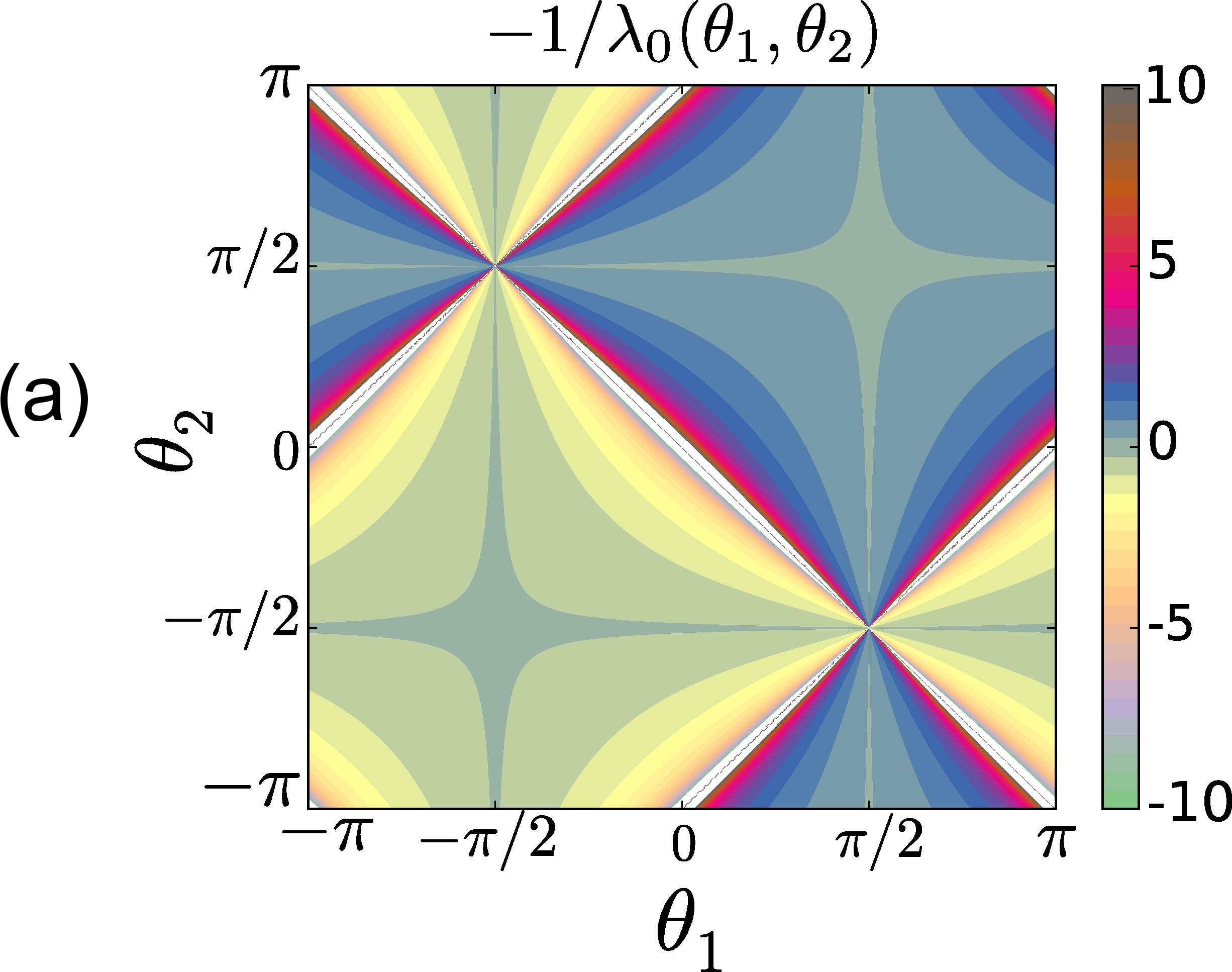}
\includegraphics[width=0.37\textwidth]{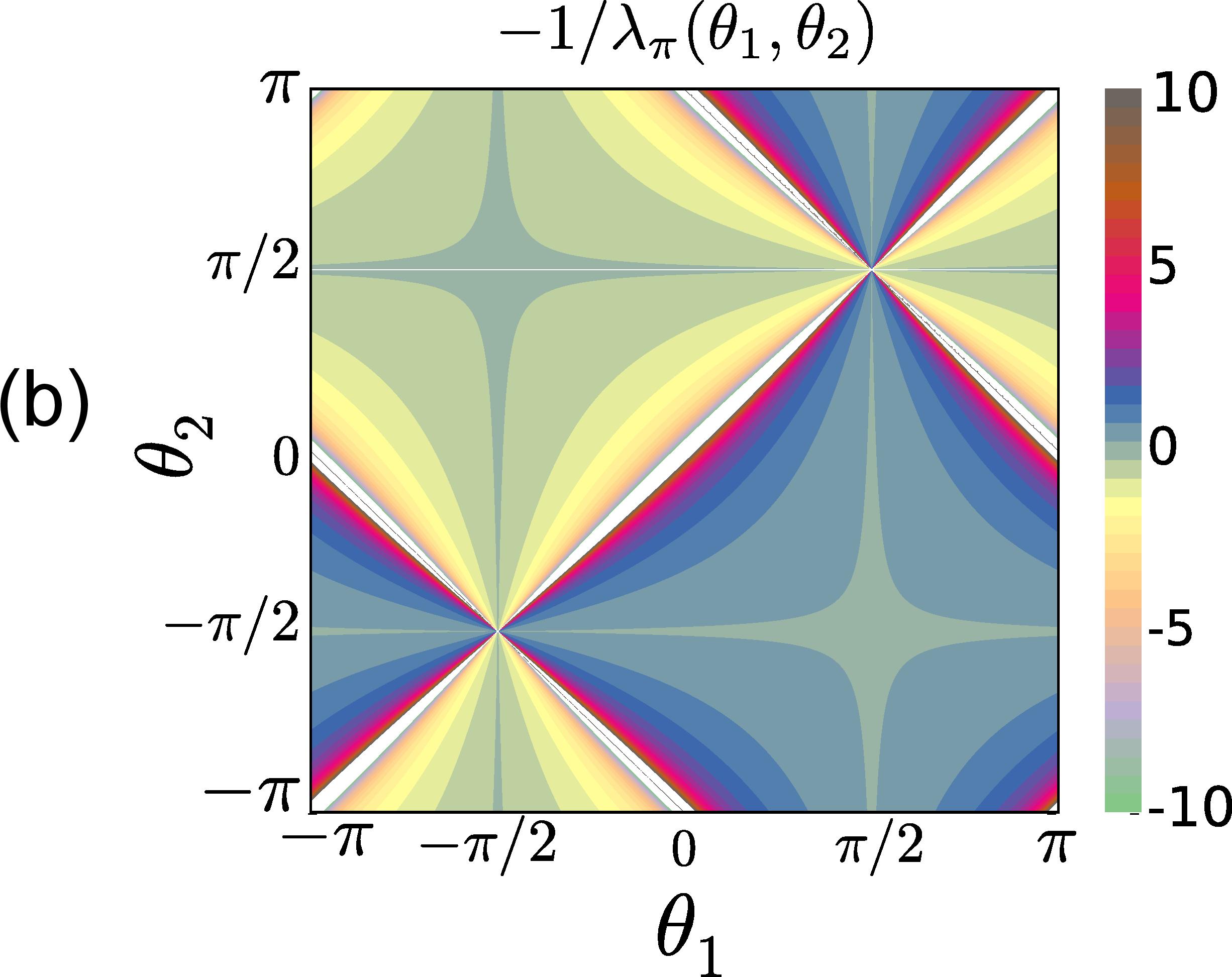}
\caption{Color online) The negative inverse of the Lyapunov exponents,
  $-1/\lambda_0$ and $-1/\lambda_\pi$, as functions of the rotation
  angles $\theta_1$ and $\theta_2$, as described in
  Eqs.~\eqref{eq:Lyapunov_exponents}. Along the white lines,
  $\lambda_E \ll 1$, thus these points are on the topological phase
  boundary. Comparison with Fig.~\ref{fig:phasemap} shows that the
  signs of the Lyapunov exponents give the correct topological
  invariants as per Eq.~\eqref{eq:phases_from_Lyapunov}.}
\label{fig:Lyapunov_exponents}
\end{figure}

The reflection and transmission amplitudes of the whole system at the
relevant quasienergies are obtained from
Eqs.~\eqref{eq:reflection_from_realspace_smatrix} using the Lyapunov
exponents as 
\begin{subequations}
\begin{align}
\label{eq:reflection_lambda}
r(E) &= -\tanh(L \lambda_E); \\
\label{eq:transmission_lambda}
t(E) &= 1/\cosh(L \lambda_E) \qquad \qquad (E = 0, \pi).
\end{align}
\end{subequations}
In the thermodynamical limit of $L\to \infty$, if the Lyapunov
exponents are nonzero, the effective Hamiltonian of the quantum walk
is an insulator, and $\abs{r(E)}\to 1$. In that case, the topological
invariants are obtained from
Eq.~\eqref{eq:invariants_from_scattering},
\begin{equation}
\label{eq:phases_from_Lyapunov}
\nu_E = -\frac{1}{2}\text{sign}(\lambda_E).
\end{equation}
For a translation invariant system, the Lyapunov exponent is the same
as the additive parameter of a single site, $\lambda_E = \lambda_x(E)$
as per Eq.~\eqref{eq:Lyapunov_exponents}.  The topological invariants
as a function of the global parameters $\theta_1$ and $\theta_2$, as
read off from the plots of Fig.~\ref{fig:Lyapunov_exponents}, agree
with the previously known results of Fig.~\ref{fig:phasemap}.

While the signs of the Lyapunov exponents give us the two topological
invariants, their absolute values tell us about the degree of
localization of the states with quasienergies $E=0$ and $\pi$.  To see
this we define the quasienergy dependent \emph{localization length}:
\begin{equation}
\xi(E) = -\frac{2L}{\llog(|t(E)|^2/4)}
\end{equation} 
where it is assumed that the length of the scattering region is
$L\to\infty$.  From Eq.~\eqref{eq:transmission_lambda} it follows,
that for large $L$, the transmission amplitude at quasienergy $E=0$ or
$\pi$ can be approximated by $t(E) \approx
2e^{-L|\lambda_E|}$. The localization length at these
energies is therefore related to the Lyapunov exponent:
\begin{equation}
\label{eq:loclength}
	\xi(E) = 1 / |\lambda_E| \qquad (E=0,\pi)
\end{equation} 

Transmission at quasienergies $E=0,\pi$ decays exponentially with the
system size whenever $\lambda_E \neq 0$. This shows that
in any of the topological phases the quantum walk is insulating at
both $E=0$ and $E=\pi$. On the phase boundary however, where
$\lambda_E$ changes sign, $\xi(E)$ diverges and the walk
is no longer insulating at one of these energies.

\begin{figure}[h!]
 \centering
 \includegraphics[width=0.95\columnwidth]{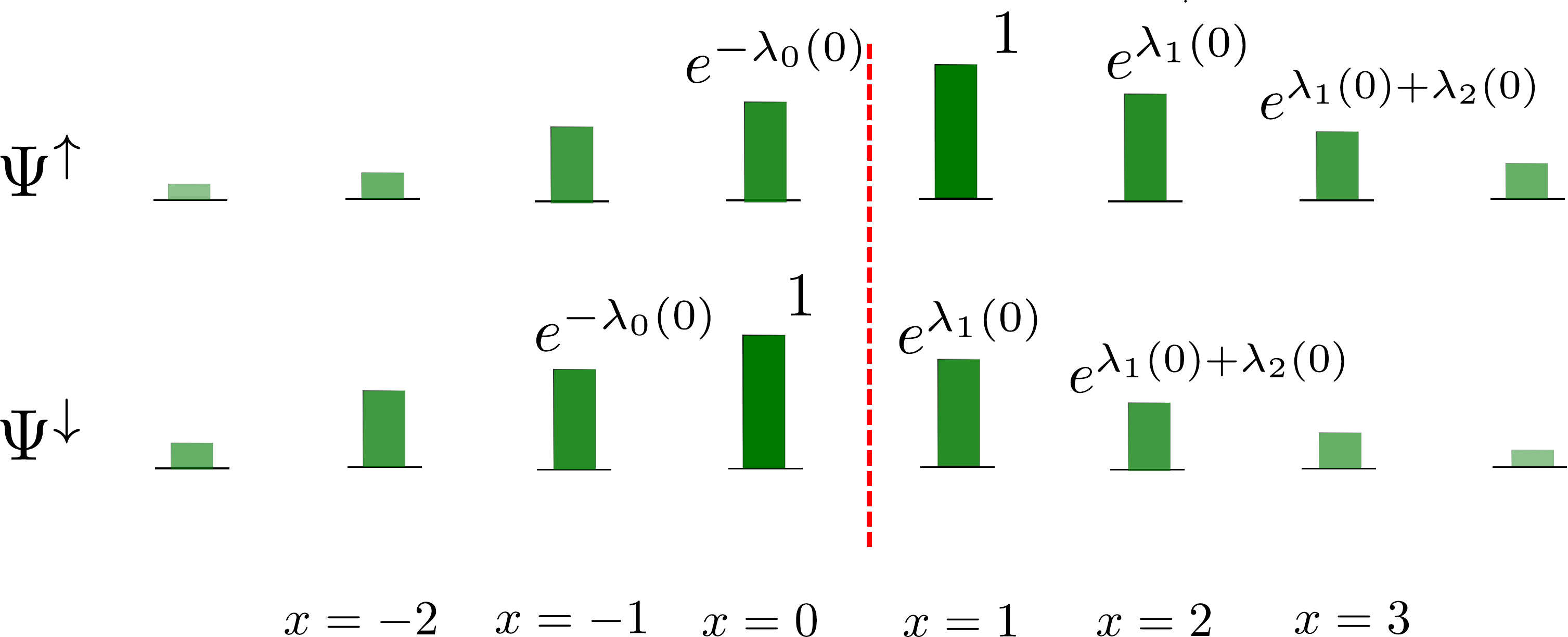}
 \caption{Example for the wavefunction of a zero-quasienergy edge
   state near the interface of two disordered bulks with differing
   topological invariants ($x<0: \lambda_0>0$ and $x>0: \lambda_0<0$).
   We chose $(\Psi_1^\uu, \Psi_0^\dd)$ to be $(1,1)$, thus the two
   components of the spinor wavefunction (upper row, lower row) are
   real and positive, and the wavefunction is not normalized to 1.
The components of the wavefunction simply acquire factors of
$e^{\pm|\lambda_E^x}|$ as we move away from the interface, so that the
edge state decays exponentially into both bulks.}
 \label{fig:edgestates_decay}
 \end{figure}

At $E=0$ and $\pi$, the localization length $\xi(E)$ also defines the
characteristic size of the edge states. To see this, suppose that we
have an interface between two bulks characterized by different
Lyapunov exponents: $\lambda_E^r$ for $x \geq 1$, and $\lambda_E^l$
for $x \leq 0$. We want to study the existence of zero energy edge
states localized near this interface. If we chose the components of
the wavefunction $(\Psi_1^\uu, \Psi_0^\dd)$ at the interface to be an
eigenstate of $\sigma_x$ with eigenvalue $\pm 1$, then according to
Eq.~\eqref{eq:tmatrix_parametrization}, at $L \gg 1$ distance from the
interface, the wavefunction will be proportional to $\exp(\pm
L\lambda_0^r)$ in the right bulk and $\exp(\mp L\lambda_0^l)$ in the
left bulk, as shown in Fig.~\ref{fig:edgestates_decay}. Since $\xi(0)
= 1/\lambda_0$, this indeed means that the edge state decays into the
bulk with the characteristic length $\xi(0)$. Similar argument holds
for $\pi$ energy edge states with an appropriate choice of boundary
conditions at the interface.  The above argument also shows that the
bulk-boundary correspondence holds for the topological invariants
defined in Eq.~\eqref{eq:phases_from_Lyapunov}, since the only way to
create normalized edge states is if the sign of the Lyapunov exponents
is different in the two bulks.

At a phase transition between different topological phases, at least
one of the $\lambda_E$ changes its sign. Thus, as we approach the
phase boundary, the corresponding characteristic length scale diverges
and the states with quasienergy $E=0$ or $\pi$ that were previously
localized become delocalized throughout the whole system, i.e., their
support will scale with the system size $L$, which in turn leads to
high values of transmission probability at these energies. This
delocalized behaviour leads to a subdiffusive propagation of a walker
started from the origin, as way previously noted in the case of the
simple walk.\cite{obuse_delocalization}

In the limit of infinite system length, we can write down exact
formulas for the Lyapunov exponents even in the disordered case. As
already stated, by disorder we mean a probability measure
$\mu(\theta_1,\theta_2)$ given on the parameter space, according to
which the angles $\theta_1$ and $\theta_2$ are chosen at each site. As
we increase the length of the system, the sum $\sum_x \lambda_x(E)$
performs a random walk, with the coordinate $x$ playing the role of
the time variable and $\lambda_x(E)$ playing the role of the distance
covered in the $x$-th step. The two topological phases correspond to
the cases when the random walk of the Lyapunov exponents drifts to the
plus/minus infinity, where $\lambda_E$ can be thought of as a
time-averaged drift velocity. For a system tuned to the topological
phase boundary, the left and right drift terms cancel out and the
random walk remains centered around the origin. Since the Lyapunov
exponents are self-averaging, in the $L\rightarrow\infty$ limit, the
''time average'' $\lambda_E$ becomes the ensemble average,
\begin{equation}
\label{eq:lambda_avg}
\lambda_E 
= \int \mathrm{d}\mu(\theta_1,\theta_2) \lambda_x(E,\theta_1,\theta_2)
\end{equation}
According to Eq.~\eqref{eq:phases_from_Lyapunov}, the sign of the above
integral gives the topological invariant, while its absolute value
defines the localization length as seen in Eq.~\eqref{eq:loclength}
This means that Eq.~\eqref{eq:lambda_avg}, together with the
definition of the Lyapunov exponents,
Eq.~\eqref{eq:Lyapunov_exponents}, enables us to calculate the exact
topological invariants and localization lengths for any disorder given
in the form of $\mu(\theta_1,\theta_2)$. This is the main result of
this paper.

\section{Split-step walk with uniform disorder in the rotation angles}
\label{sec:uniform}

\begin{figure}[h!]
 \centering
 \includegraphics[width=0.35\textwidth]{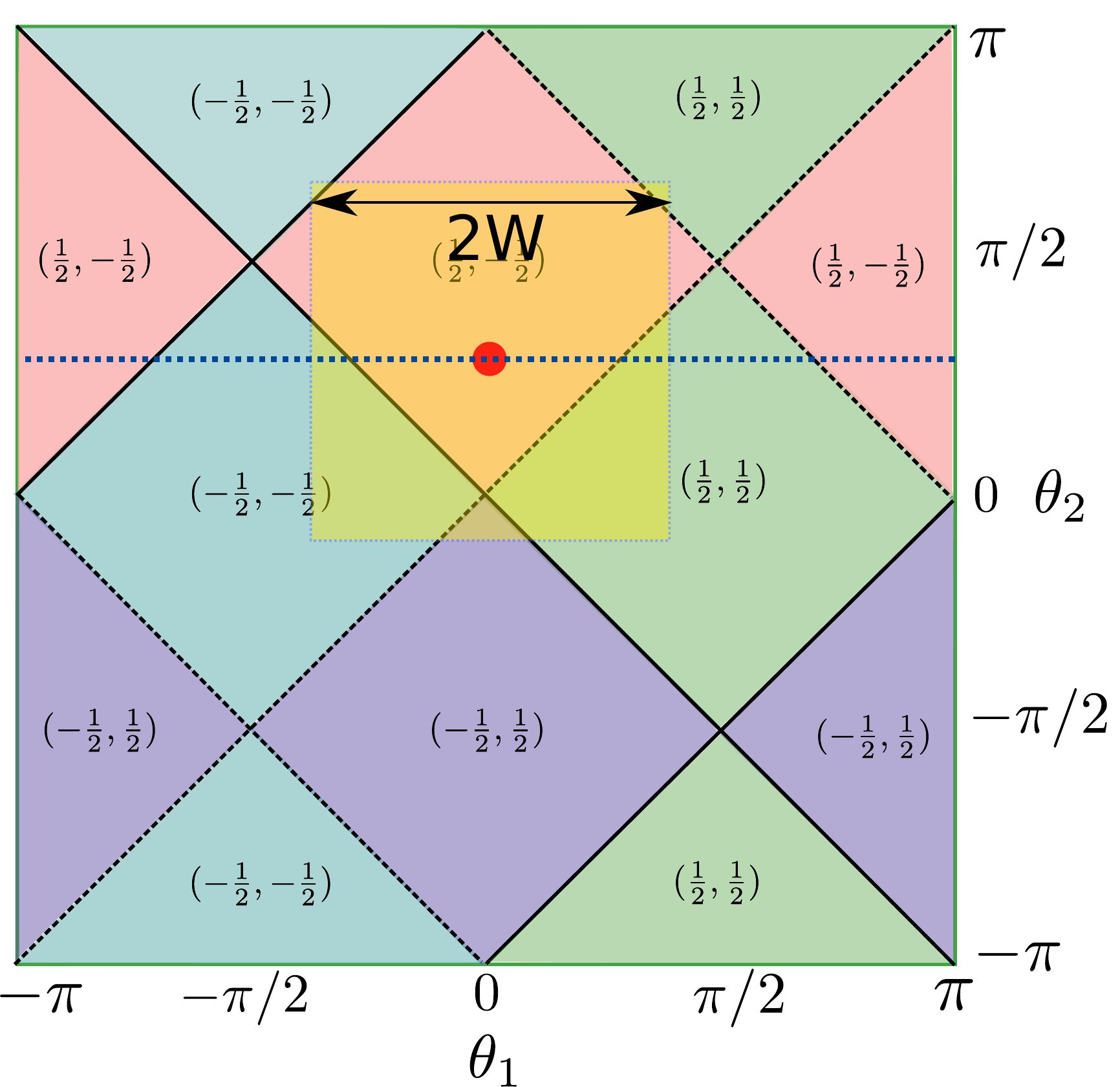}
 \caption{(Color online) The split-step walk with uniform disorder in
   the rotation angles. The parameters are taken from the yellow
   square of size $2W\times2W$, with uniform probability. In 
   Fig.~\ref{fig:qw_uniform_maps} we move the middle of the
   distribution along the dashed line and vary the value of
   $W$ from $0$ to $\pi$.}
 \label{fig:phasemap_uniform_disorder}
 \end{figure}

As an illustration of the ideas developed in the previous Section we
calculate the topological phase map of the split-step walk with
uniform disorder.  We take the rotation angles $\theta_1$ and
$\theta_2$ randomly and independently from a box distribution of width
$W$ centered around mean values $\expect{\theta_1},
\expect{\theta_2}$.  The corresponding probability density function,
\begin{multline}
\label{eq:uniform_def}
\mu(\theta_1,\theta_2) = \frac{1}{4W^2}(\Theta(\langle\theta_1\rangle + 
W - \theta_1) \cdot \Theta(\theta_1 - \langle\theta_1\rangle + W)) \\
\cdot(\Theta(\langle\theta_2\rangle + W - \theta_2) \cdot 
\Theta(\theta_2 - \langle\theta_2\rangle + W)),
\end{multline}
is illustrated in Fig.~\ref{fig:phasemap_uniform_disorder}.
(For another illustrative example, that of binary
disorder, see App.~\ref{sec:app_binary}.)

\begin{figure}[h!]
\centerline{%
\includegraphics[width=0.53\columnwidth]{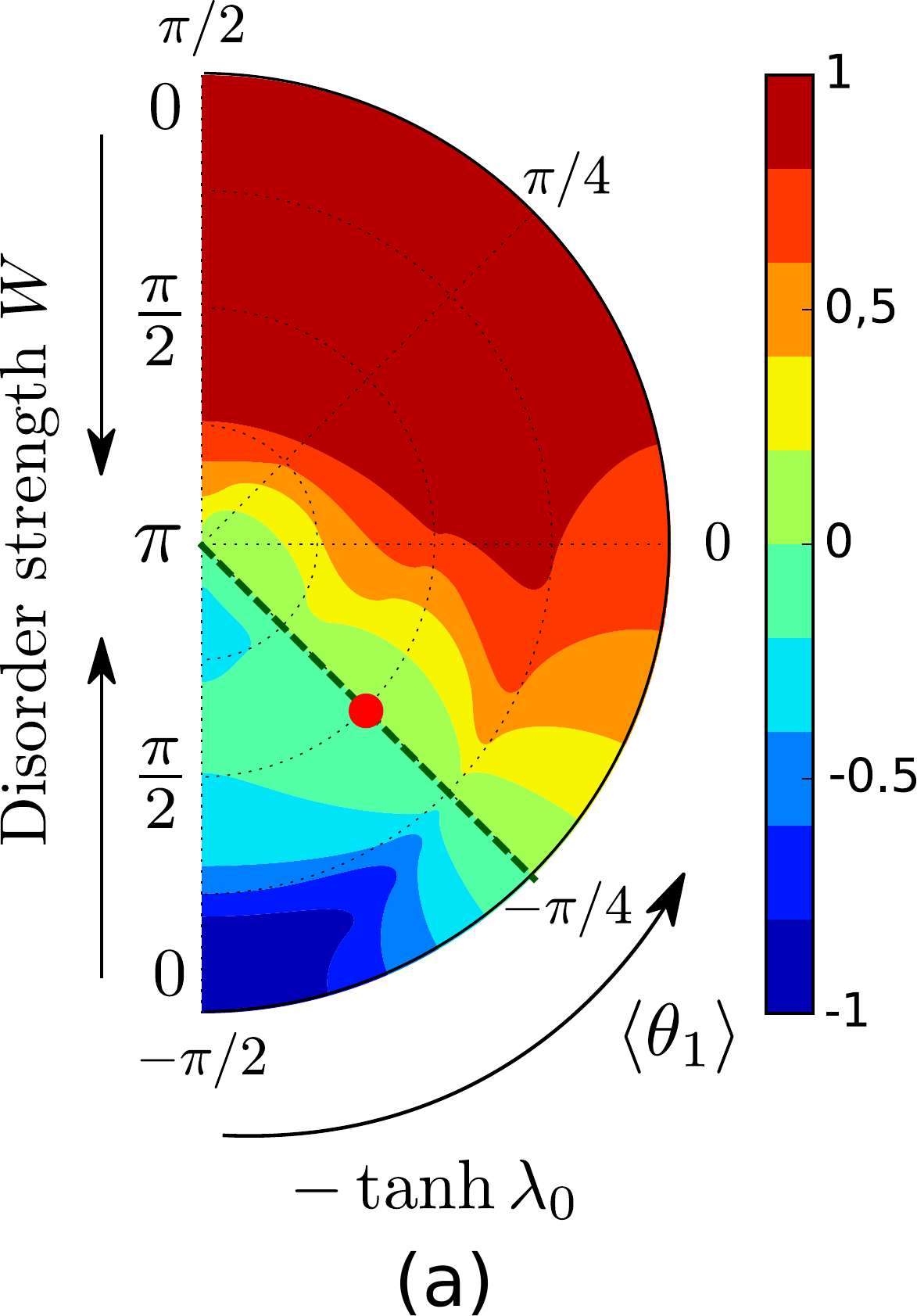}%
\includegraphics[width=0.47\columnwidth]{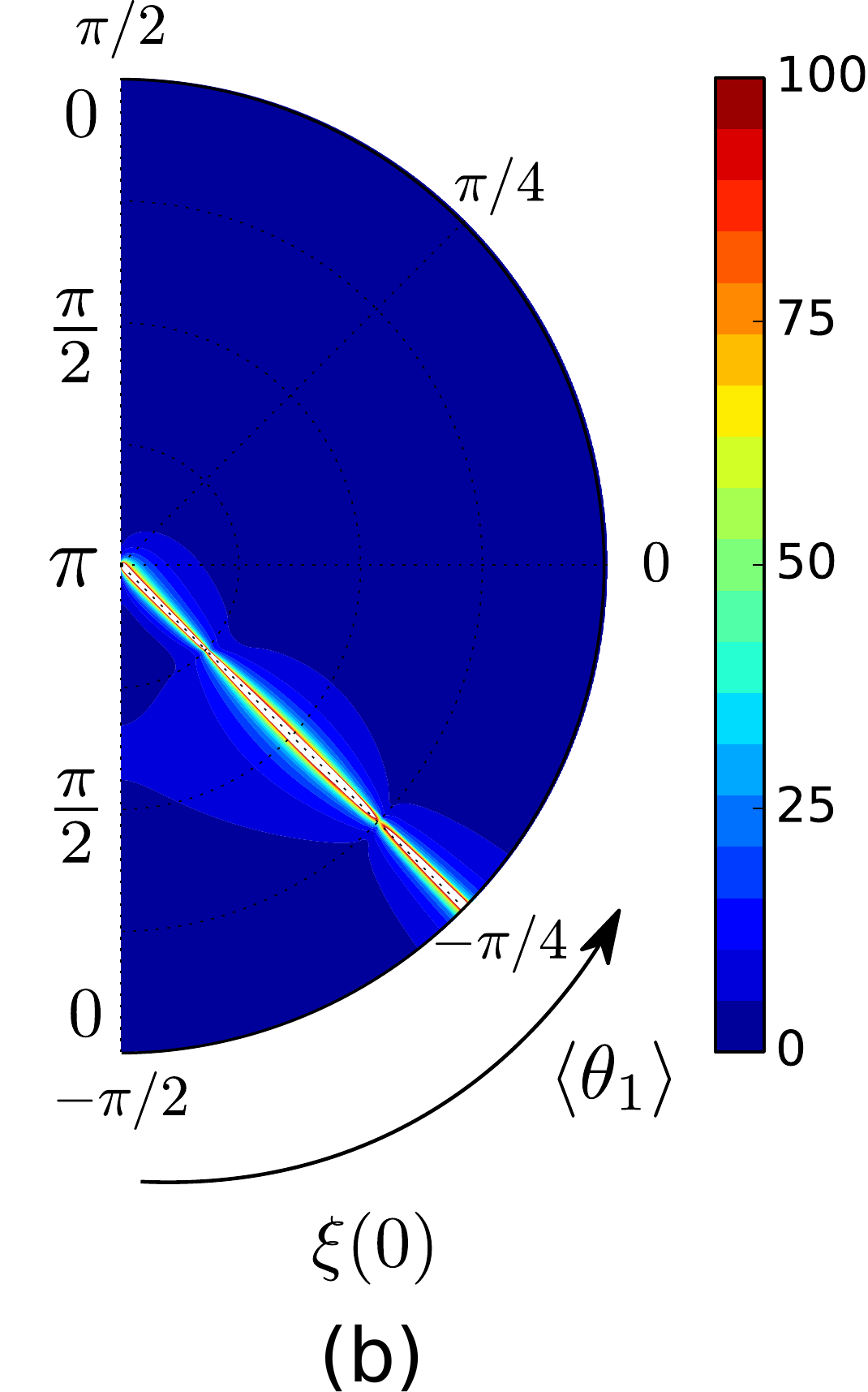}%
}%
\label{qw_uniform_maps4}
\caption{(Color online) Topological phase map of the split-step walk
  with $\langle\theta_2\rangle = \pi/4$ and uniform disorder in both
  rotation angles. The angle of the polar plot is the average
  $\langle\theta_1\rangle$ while disorder strength increases in the
  radial direction so that the middle of the circle corresponds to the
  strong disorder limit (see text). (a) Average reflection ampitude
  for a system consisting of a single site, $-\tanh{\lambda_0}$. The
  dashed line separates the two topological phases with $\nu_0 = -1$
  and $\nu_0 = +1$, the (red) dot indicating the values of the
  parameters used in Fig. \ref{fig:loclength_divergence}.  (b)
  Localization length, computed for a system of infinite size. Along
  the phase boundary, the localization length $\xi(0)$ diverges.}
\label{fig:qw_uniform_maps}
\end{figure}

In Fig.~\ref{fig:qw_uniform_maps}, we show the effects of disorder on
the zero-quasienergy Lyapunov exponent, more precisely, on the average
reflection amplitude for a system of size 1, and on the localization
length.  For simplicity, here we set the mean second rotation angle
to $\expect{\theta_2} = \pi/4$, and we vary the mean of the first
rotation angle, $\expect{\theta_1}$ from $-\pi/2$ to $\pi/2$ (along
the dashed line in Fig.~\ref{fig:phasemap_uniform_disorder}), as well
as the disorder strength $W$. At each point in this phase map we
numerically integrate Eq.~\eqref{eq:lambda_avg} to obtain the Lyapunov
exponent $\lambda_0$, and from it, the mean reflection amplitude
$-\text{tanh} \,\lambda_0$ as well as the localization length at 0
quasienergy $\xi_0$.
We find that the boundary between the two topological phases, where
$\lambda_0=0$ (dashed line in Fig.~\ref{fig:qw_uniform_maps}), is at
$\expect{\theta_1}=-\pi/4$, independent of the disorder $W$.  Along
this line, the system is critical, and the localization length
diverges at quasienergy $E=0$.

\subsection{Disorder-induced delocalization} 

\begin{figure}[h!]
 \centering
 \includegraphics[width=0.9\columnwidth]{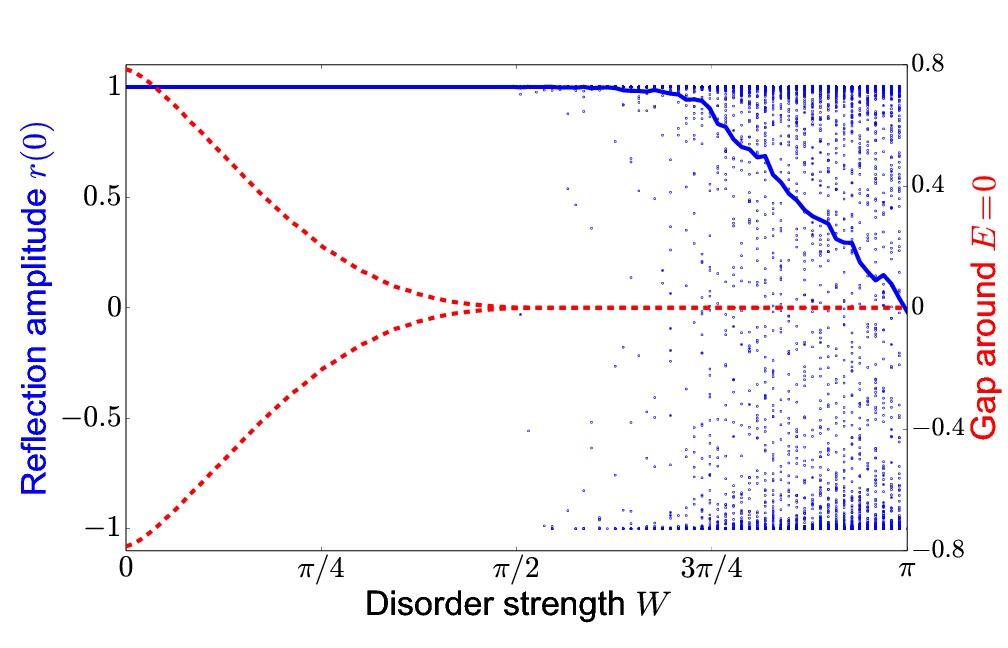}
 \caption{(Color online) Topological phase change as a result of
   increasing disorder, in a system of 200 sites, averaged over 1000
   disorder realizations. Two branches of the dashed (red) curve: the
   highest/lowest quasienergy in the lower/upper band.  The distance
   between the branches -- the gap around zero quasienergy --
   gradually closes as disorder increases.  Solid (blue) curve: the
   average zero quasienergy reflection amplitude. As the disorder
   increases, this remains quantized at $r(0) = 1$ as necessitated by
   Eq.~\eqref{eq:invariants_from_scattering} long after the gap
   closes. As the disorder strength is increased even further,
   $\langle r(0)\rangle $ begins to change and eventually at $W = \pi$
   it reaches $0$, signifying a phase transition point. 
   The (blue) points represent $r(0)$ in the 1000 individual
   realizations whose average is the solid blue curve.}
 \label{fig:transition_at_strong_disorder}
 \end{figure}

The system undergoes a disorder-induced delocalization transition in
the \emph{strong disorder limit} of $W=\pi$, where the rotation angles
are taken from the whole parameter space with equal probabilites. At
this limit, which corresponds to the central point in
Fig.~\ref{fig:qw_uniform_maps}, we find analytically $\lambda_0 =
\lambda_\pi = 0$. Thus, starting from a clean system, by increasing
the strength of the disorder, we eventually reach a topological phase
transition point, as shown in
Fig.~\ref{fig:transition_at_strong_disorder}. This is accompanied by a
delocalization of states with energies $E = 0$ and $E = \pi$ as
described in the previous section. Thus, while at small values of the
disorder strength, the quantum walk is Anderson localized at all
energies (we confirmed this separately by numerically calculating the
localization lengths at other energies), strong disorder induces
delocalization at the specific energies protected by the symmetries of
the walk. This is similar to the case of the SSH
model\cite{mondragon2013topological}. We note that the qualitative
difference with respect to Ref.~\onlinecite{mondragon2013topological},
where in the infinite-disorder limit a transition to a trivial
localized phase was found, in we checked that using different disorder
for $\theta_1$ and $\theta_2$ leads to a transition from one
topological phase to another, rather than delocalization.

\subsection{No Anderson localization on the critical line} 

There are special cases for the split-step walk where disorder does
not induce Anderson localization: if at some quasienergy the average
Lyapunov exponent, defined by Eq.~\eqref{eq:lambda_avg}, vanishes.
Uniform disorder in the rotation angles realizes such a special case
if the quantum walk is on average on a phase boundary, i.e., if
\begin{align}
\label{eq:condition_theta_critical}
\expect{\theta_1} = \pm\expect{\theta_2} + n \pi
\end{align}
for some $n\in\mathbb{Z}$, as seen in
Fig.~\ref{fig:phasemap_uniform_disorder}. 
In these cases, the localization length has to diverge at some
critical quasienergy. As we show below, both the critical quasienergy
and the shape of the divergence can be explained by a mapping to the
disordered simple quantum walk.

\begin{figure}[h!]
\centering
\includegraphics[width=0.49\columnwidth]{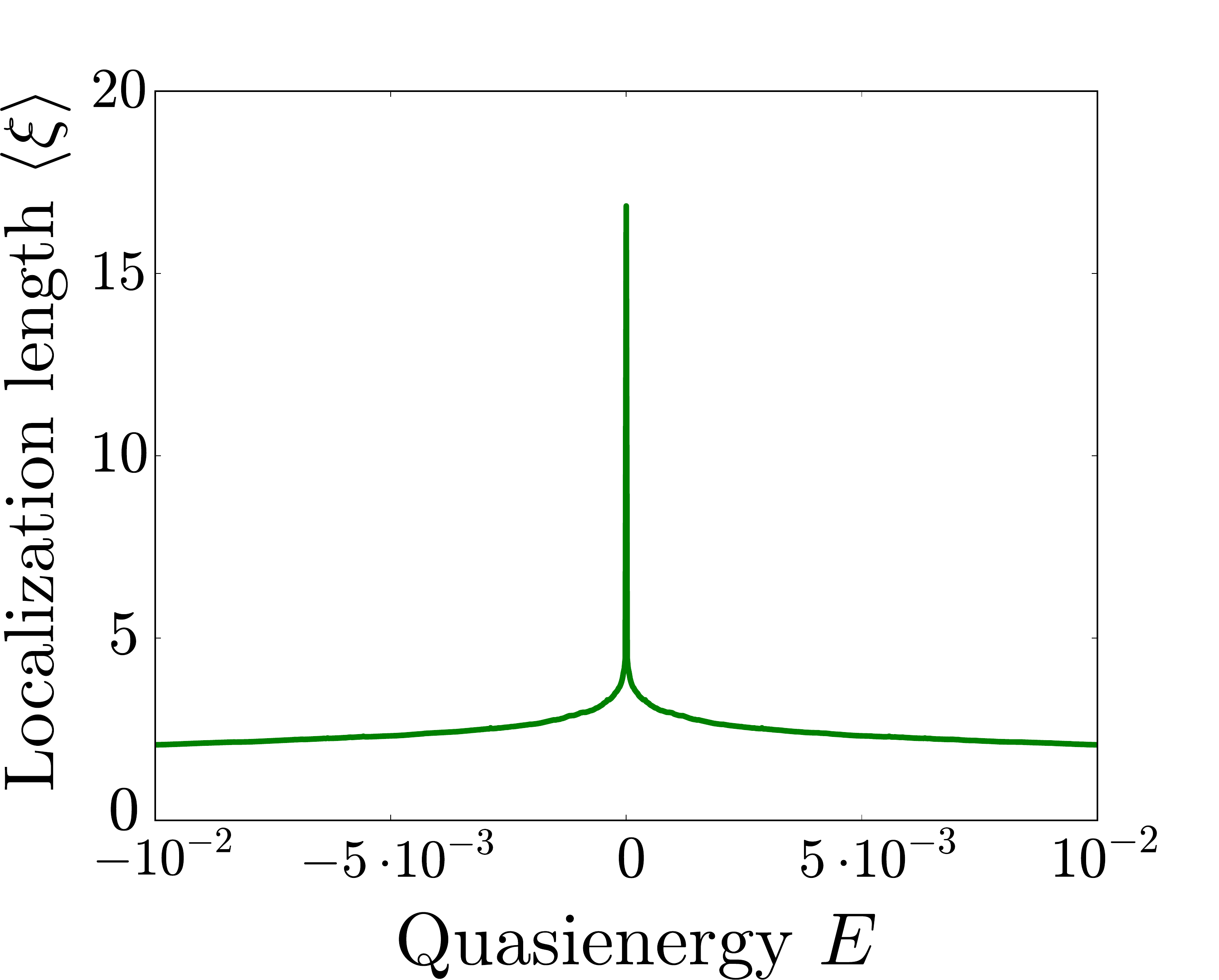}%
\includegraphics[width=0.49\columnwidth]{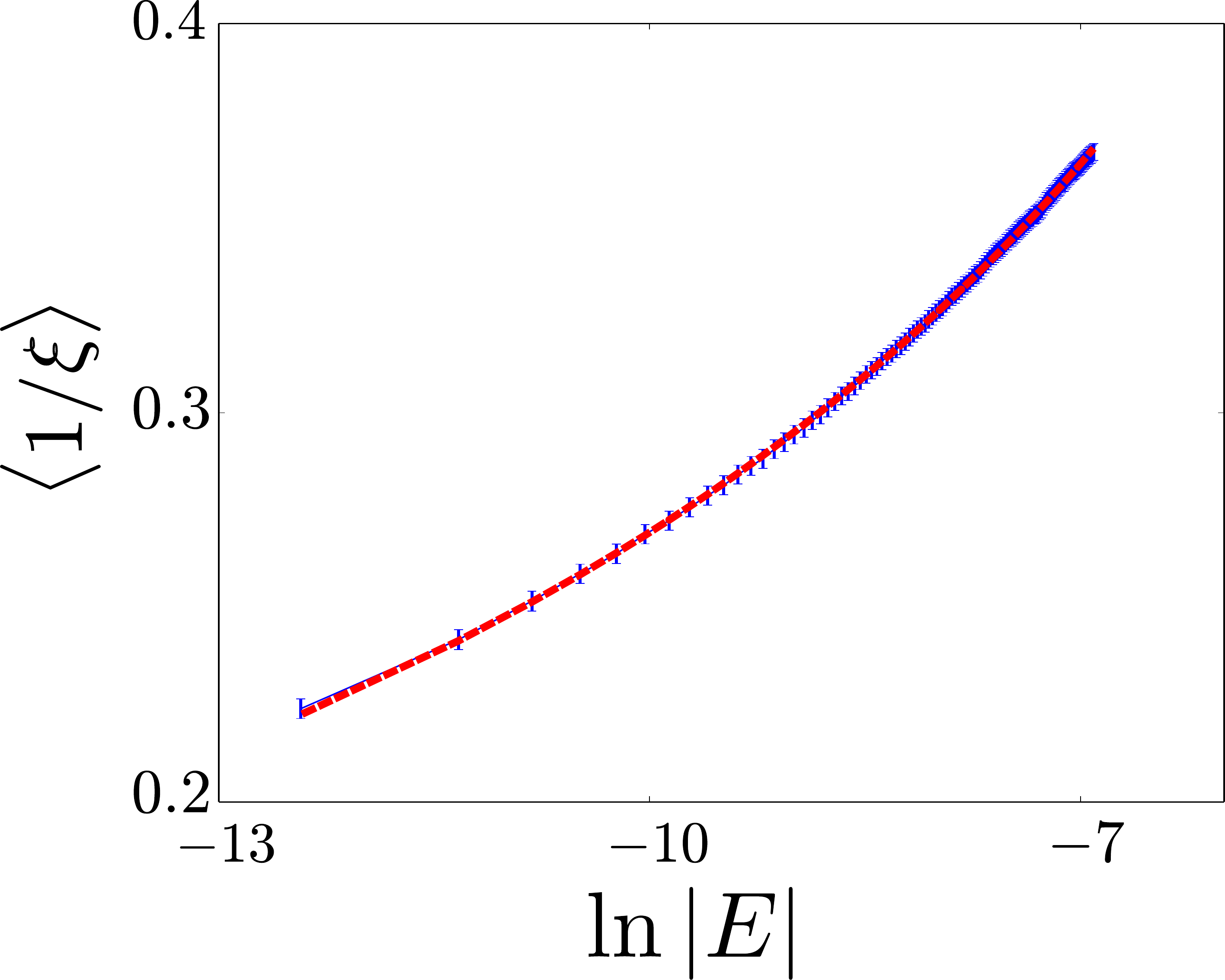}
\caption{(Color online) Divergence of the localizaton length $\xi(E)$ as a function
  of quasienergy $E$ at a critical point with mean angles
  $\expect\theta_1 = -\expect\theta_2 = -\pi/4$ and disorder $W =
  \pi/2$ (see Fig. \ref{fig:qw_uniform_maps}/(b)). (a) The average
  localization length diverges at $E=0$. (b) The scaling of the
  localization length near the critical energy. The dotted line
  indicates the curve \eqref{eq:loclength_scaling} fitted to the
  numerical data, which gives $\tau = 0.216$ and $\xi_0 = 0.322$. We
  used a system of 100 and 400 sites for figures (a) and (b)
  respectively. In both cases we averaged over 100 disorder
  realizations.}
\label{fig:loclength_divergence}
\end{figure}

Obuse and Kawakami\cite{obuse_delocalization} have found that the
simple walk, defined in Eq.~\eqref{eq:U_simple_def}, with uniform
disorder in the rotation angle $\theta$, does not undergo Anderson
localization, no matter how strong the disorder is.  
A key part of their explanation is that apart
from the chiral and particle-hole symmetries, the simple walk posesses
a sublattice symmetry, 
\begin{align}
\Lambda U \Lambda &= -U \qquad \Lambda = 
\sum_{\text{x even}} \ket{x}\bra{x} - \sum_{\text{x odd}} \ket{x}\bra{x},
\label{eq:sublattice_symmetry_def}
\end{align}
because the walker can only hop from even to odd sites or vice versa
in one step.  Using this extra symmetry, they argued (for more
details, see Ref.~\onlinecite{zhao2015disordered}), that Anderson
localization is avoided because the localization length $\xi$ of the
simple quantum walk diverges at $E=\pm\pi/2$, scaling as
\begin{equation}
\label{eq:loclength_scaling}
\xi(E) = \xi_0 |\log(\delta E \tau)|,
\end{equation}
where $\delta E = E - E_\text{crit}$ is the distance from the critical
quasienergy $E_\text{crit}=\pm\pi/2$, and $\tau$ is the mean free
time.  As illustrated in Fig.~\ref{fig:loclength_divergence}, the same
effect occurs in split-step walks, if
Eq.~\eqref{eq:condition_theta_critical} is fulfilled.  We show this
explicitly in the rest of this section.

Using the sublattice symmetry property of
Eq.~\eqref{eq:sublattice_symmetry_def}, a split-step quantum walk can
be understood as a doubled sequence of the simple quantum walk.
Consider
\begin{multline}
\label{eq:simple_in_splitstep}
\bra{x',s'} [S_\downarrow R(\theta_2) S_\uparrow R(\theta_1)]^t \ket{x,s} = \\ 
\bra{2x',s'} [S R(\theta') S R(\theta')]^t \ket{2x,s},
\end{multline}
with the position dependent rotation angles $\theta'(x)$ defined by
the relations
\begin{align}
\theta'(2x) &= \theta_1(x);\\
\theta'(2x-1) &= \theta_2(x).
\end{align}
This mapping is closely related to that used in
Ref.~\onlinecite{asboth2015edge} to realize a split-step quantum walk
as a periodically driven Hamiltonian. 

The mapping of Eq.~\eqref{eq:simple_in_splitstep} shows that if
$\expect{\theta_1}=\expect{\theta_2}$, the correlation length of the
split-step walk has to diverge at quasienergy $E_\text{crit}=\pi$
following the scaling of Eq.~\eqref{eq:loclength_scaling}. By
Eq.~\eqref{eq:simple_in_splitstep}, two timesteps of a simple quantum
walk with angle $\theta$ can be understood as a single timestep of a
split-step quantum walk with $\theta_1 =\theta_2 = \theta$.  The
introduction of uncorrelated box disorder in the rotation angle
$\theta$ of the simple quantum walk translates to uniform disorder in
the angles $\theta_1$ and $\theta_2$ of the quantum walk, with
$\expect{\theta_1}=\expect{\theta_2}$. The doubling of the timestep
moves the critical quasienergy from $\pm \pi/2$ to $\pm \pi$.

A mapping between different split-step quantum walks shows that if
$\expect{\theta_1}=\expect{\theta_2}\pm\pi$, the divergence of the
localization length follows the same functional form of
Eq.~\eqref{eq:loclength_scaling}, with $E_\text{crit}=0$.  
Changing $\theta_j(x) \to \theta_j(x)\pm\pi$ for all $x$, for either
$j=1$ or $j=2$, results in an overall factor of $-1$ for the timestep
operator $U$. This shifts the divergence of the localization length of
the quantum walk by $\pi$, to $E_\text{crit}=0$. 

A mapping between the transfer matrices of different split-step
quantum walks shows that if
$\expect{\theta_1}=-\expect{\theta_2}+n\pi$, the localization length
diverges according to Eq.~\eqref{eq:loclength_scaling}, with shifted
$E_\text{crit}$. According to Eq.~\eqref{eq:tmatrix}, changing either
$\theta_1(x)$ to $-\theta_1(x)$ or $\theta_2(x)$ to $-\theta_2(x)$
changes the transfer matrix $M_x$ at $x$ to $M_x'$, with
\begin{align}
M_x'(E+\pi) = -
\begin{pmatrix} i & 0 \\ 0 & 1 \end{pmatrix}
M_x(E)
\begin{pmatrix} -i & 0 \\ 0 & 1 \end{pmatrix}. 
\end{align}
The transfer matrix $M_x'(E+\pi)$ is a unitary transform of $M_x(E)$,
up to the unimportant factor of $-1$. Since the transformation is
independent of $x$, the Lyapunov exponents at $E$ of the original walk
are equal to those of the transformed walk at $E+\pi$. This shows that
if $\expect{\theta_1}=-\expect{\theta_2}$, the localization length
diverges at $E_\text{crit}=0$, while if
$\expect{\theta_1}=-\expect{\theta_2}\pm\pi$, it diverges at
$E_\text{crit}=\pi$, following the scaling of
Eq.~\eqref{eq:loclength_scaling}.


Our results also shed new light on the absence of Anderson
localization in the simple quantum walk\cite{obuse_delocalization}.
As we have shown, the simple quantum walk with disorder in the
rotation angles is not Anderson localized because it constitutes a
disordered split-step quantum walk tuned to a topological phase
transition point.

\section{Split-step walk with phase disorder}
\label{sec:phase_disorder}

We now consider the phase disordered split-step walk, introduced in
Eq.~\eqref{eq:U_p_def}. Since phase disorder breaks both chiral and
particle-hole symmetry, we expect that it induces Anderson
localization at all quasienergies. There is a way, however, to
add phase disorder to the split-step walk and keep chiral symmetry: in
that case, we expect to see localization-delocalization transitions as
with angle disorder in the previous Section. We discuss both types of
phase disorder, and illustrate our results by numerical examples that
can be compared directly with those on angle disorder of the previous
Section.

The simplest way to introduce phase disorder is to multiply the
wavefunction of the walker at the end of each timestep by a position-
and spin-dependent phase factor $\phi(x,s)$, chosen randomly and
independently at each site, as defined in Eq.~\eqref{eq:U_p_def}.  For
the examples in this section we used an extra restriction of
$\phi(x,\uparrow) = -\phi(x,\downarrow)$, 
whereby the phase operator reads
\begin{equation}
P(\phi) = \sum_x |x\rangle\langle x| \otimes e^{-i\phi(x)\sigma_z}, 
\end{equation}
with the phase chosen from an interval $[-\Delta\phi,\Delta\phi]$ with
uniform distribution.
Just as with the more general phase disorder, due to this extra
operation, both particle-hole symmetry and chiral symmetry of the
quantum walk are broken. Thus, in the presence of phase disorder,
there are no localization-delocalization transitions.  

To highlight the role of chiral symmetry, we also consider adding
phase disorder to the split-step walk in a chiral symmetric way. 
This requires two phase operators per timestep,
\begin{equation}
\label{eq:double_phase_disorder}
U'(\phi,\theta_1,\theta_2) = P(\phi) S_\dd R(\theta_2) 
S_\uu P(\phi) R(\theta_1).
\end{equation}
The second phase operation restores chiral symmetry since $\sigma_x
P(\phi) \sigma_x = P(\phi)^{-1}$.  Repeating the calculation of the
real-space transfer matrix, Eq.~\eqref{eq:tmatrix}, with $R(\theta_1)$
replaced by $P(\phi) R(\theta_1) P(\phi)$, we find that this matrix is
just multiplied by a factor of $e^{2i\phi(x)}$. For example at zero
quasienergy,
\begin{equation}
M_x(0) = e^{2i\phi(x)}e^{\lambda_x(0)\sigma_x},
\end{equation}
and similarly for $E = \pi$. The extra phase factor drops out from
both the reflection amplitude and the localization length so that the
description we gave in the previous section is unaffected. 

\begin{figure}[h!]
\centerline{%
\includegraphics[width=0.46\columnwidth]{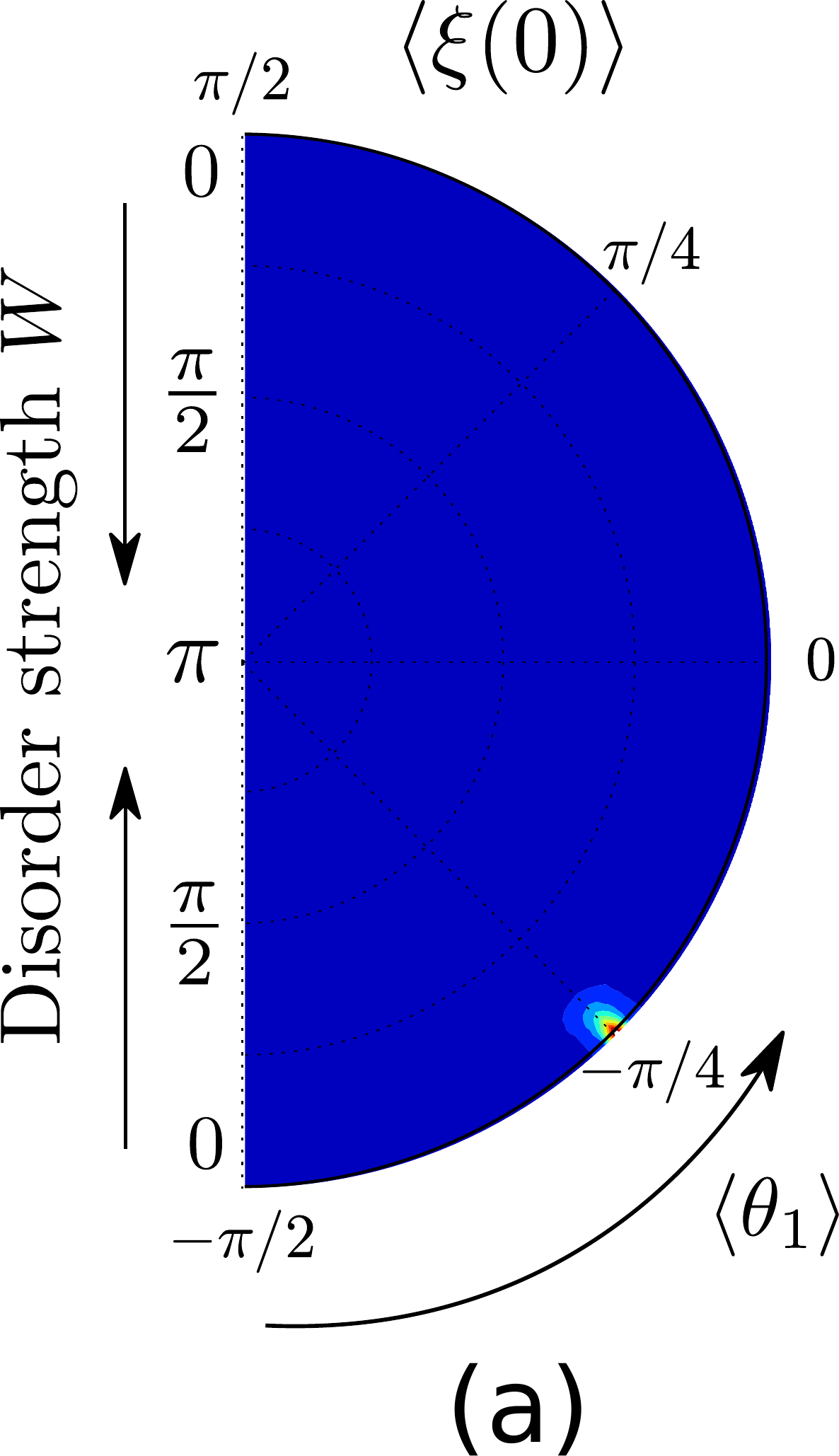}%
\includegraphics[width=0.52\columnwidth]{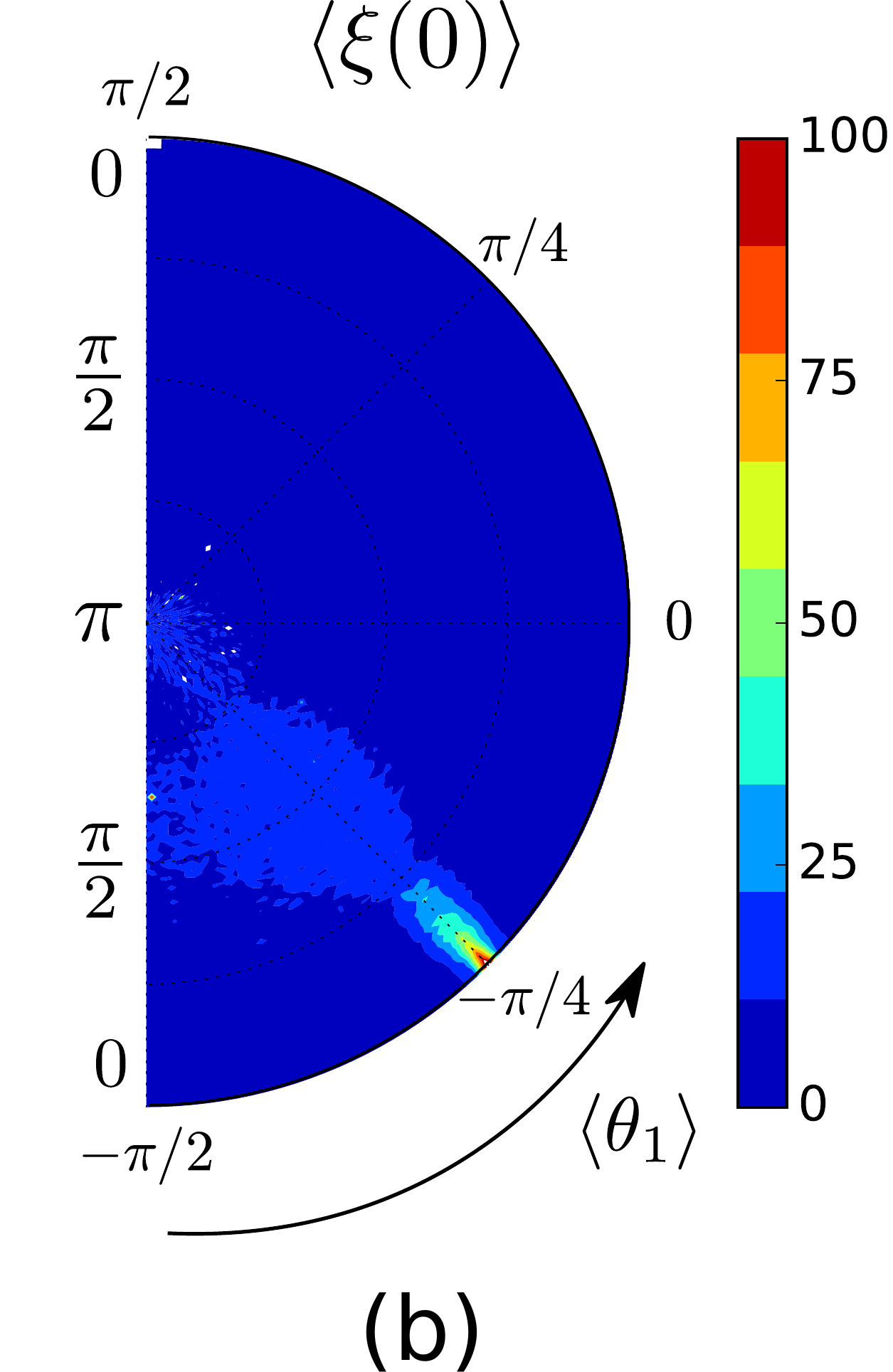}%
}%
\caption{(Color online) Average localization lengths at quasienergy $E=0$ for phase
  disordered quantum walks with equal disorder strengths for the
  rotation angles and the random phases, $\Delta\phi = W$. (a) Single
  phase shift. The timestep operator is $U = P(\phi) S_\dd R(\theta_2)
  S_\uu R(\theta_1)$ thus chiral symmetry is broken. (b) Double phase
  shift, with timestep $U' = P(\phi) S_\dd R(\theta_2) S_\uu P(\phi)
  R(\theta_1)$, exhibiting chiral symmetry. Single phase disorder
  leads to topological triviality and causes all states to be
  localized while double phase disorder restores the phase map seen in
  Fig.~\ref{fig:phasemap_uniform_disorder} with a delocalization
  transition at the topological phase border. The localization lengths
  were calculated using a system of $80$ sites and averaged over $100$
  realizations.}
\label{fig:phase_disorder}
\end{figure}

To show the effects of phase disorder numerically, we have calculated
the localization lengths $\xi$ for a range of angle disorders and
phase disorders, shown in Fig.~\ref{fig:phase_disorder}. For easy
comparison with the results of the previous Section,
Fig.~\ref{fig:qw_uniform_maps}, we fix $\expect{\theta_2}=\pi/4$, and
show the localization length $\xi(0)$ as a function of the mean first
rotation angle $\expect{\theta_1}$ and of disorder, introduced in
equal measures to both rotation angles and to the phases $\phi$,
setting $W=\Delta\phi$. Thus, the perimeter of the plot of
Fig.~\ref{fig:phase_disorder}, with $W=0$, corresponds to the
perimeter of the plots of Fig.~\ref{fig:qw_uniform_maps}b). As
expected, symmetry breaking phase disorder,
Fig.~\ref{fig:phase_disorder}a), destroys the delocalization
transition and for large values of $\Delta\phi$, the states with $0$
quasienergy are localized for all parameters.  Phase disorder that
respects chiral symmetry, Fig.~\ref{fig:phase_disorder}(b), however,
has the numerically obtained localization lengths taking large values
at $\expect{\theta_1}=-\pi/4$, which is the phase boundary between
topological phases, and where the theory (shown in
Fig.~\ref{fig:qw_uniform_maps}) predicts divergences.

\begin{figure}[h!]
\centering
\includegraphics[width=0.7\columnwidth]{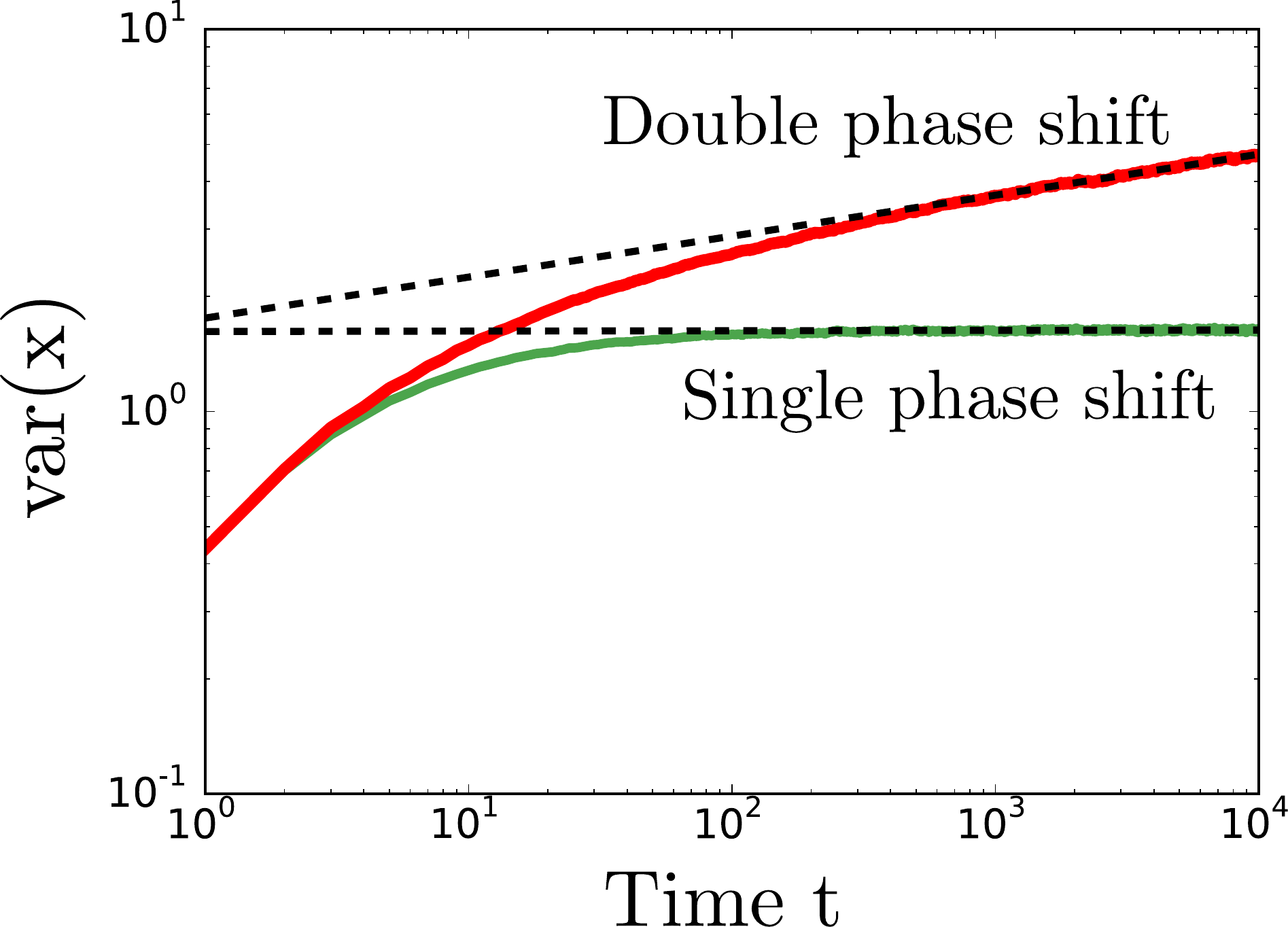}
\label{single_phase}
\caption{(Color online) Variance of the position of a quantum walker
  started from a single site as a function of time for single and
  double phase disorder, averaged over 10000 disorder realizations,
  with angle disorder strength $W = \pi$ and phase disorder strength
  $\Delta\phi = \pi/4$. For a timestep containing a single phase shift
  (green thin curve), the walker is localized and the variance tends
  to a constant, while for the chiral symmetric case defined in
  Eq.~\eqref{eq:double_phase_disorder} (red thick curve), the walker
  spreads subdiffusively. In both cases, we fitted the
  function $\text{var}(x) = a\cdot t^b$ to the data points with $t >
  1000$ and found that the exponent $b$ is $9,57\cdot 10^{-4}$ in the
  first and $0,107$ in the second case. The dashed lines show the
  fitted curves.}
\label{fig:time_evolution_phase_disorder}
\end{figure}

We also show the effects of symmetry breaking and of chiral symmetric
phase disorder on the time evolution of the split-step quantum walk
directly in Fig.~\ref{fig:time_evolution_phase_disorder}. Here we show
the position variance of a particle started from the origin after $t$
timesteps,
\begin{equation}
	\text{var}(x) = \bra{\psi(t)} x^2 \ket{\psi(t)} 
- [\bra{\psi(t)} x \ket{\psi(t)}]^2.
\end{equation} 
We used the strong disorder limit in the rotation angles, $W = \pi$,
and used phase disorder with amplitude $\Delta\phi = \pi/4$. 
For a single phase shift (symmetry breaking phase disorder), the
walker is localized and $\text{var}(x) \rightarrow \text{constant}$
for large $t$. However, in the double phase shift case, when chiral
symmetry is restored, we see the subdiffusive bevaviour discussed in
the previous Section, with the walker slowly spreading through the
system.  These results are in accordance with those from the
localization lengths.

\section{Conclusions}
\label{sec:conclusions}

We considered the effects of disorder on the localization properties
and on topological phases of the one-dimensional split-step quantum
walk. We introduced an effective numerical tool (the cloning trick),
which allowed us to calculate the scattering amplitudes for the
split-step walk and efficiently calculate the topological invariants
proposed in Ref.~\onlinecite{scattering_walk2014}. We then showed
theoretically and investigated numerically various
localization-delocalization transitions that occur whenever this
system is tuned to a critical point at a topological phase transition.
We have shown using a mapping that the subdiffusive spreading of the
simple quantum walk with angle disorder\cite{obuse_delocalization} can
be understood in this framework. We have shown that angle disorder
generically localizes the split-step quantum walk, but that complete
disorder in the rotation angles places it in a critical state with
subdiffusive instead of localized dynamics. Finally, we illustrated
the importance of symmetries on localization-delocalization through
the example of phase disorder.

It is interesting to compare our results on the one-dimensional
split-step quantum walk (1D) with those obtained in
Ref.~\onlinecite{edge2015localization} regarding the two-dimensional
split-step quantum walk (2D). In the 2D case, the topological phases
did not require any symmetry of the system, and so phase disorder did
not destroy the topological phase. Similarly to angle disorder in the
1D case, phase disorder in the 2D case was found to lead to Anderson
localization, except when the system was tuned to criticality (as in
the case of the Hadamard walk). The disorder-induced delocalization
transition however was reached in both the 1D and the 2D case by using
complete disorder in the rotation angles. In the 2D case, it was found
that angle disorder alone does not lead to Anderson localization,
possibly related to the presence of particle-hole symmetry. The effect
of particle-hole symmetric disorder remains to be studied in the 1D
case.

Our results show how the understanding of topological phases of
quantum walks can help interpret their behaviour under different types
of disorder. This could be important to identify which types of
quantum walks are practical for information processing purposes, and
which types of disorder it is crucial to supress in such applications.

\begin{acknowledgments}
We thank Andrea Alberti for useful discussions. 
This work was supported by the Hungarian Academy of Sciences
(Lend\"ulet Program, LP2011-016), and by the Hungarian Scientific
Research Fund (OTKA) under Contract No. NN109651. J.~K.~A.~also
acknowledges support from the Janos Bolyai scholarship.

\end{acknowledgments}

\bibliography{walkbib}

\appendix

\section{Split-step walk with binary disorder}
\label{sec:app_binary}

\begin{figure}[h!]
\centering
\includegraphics[width=0.8\columnwidth]{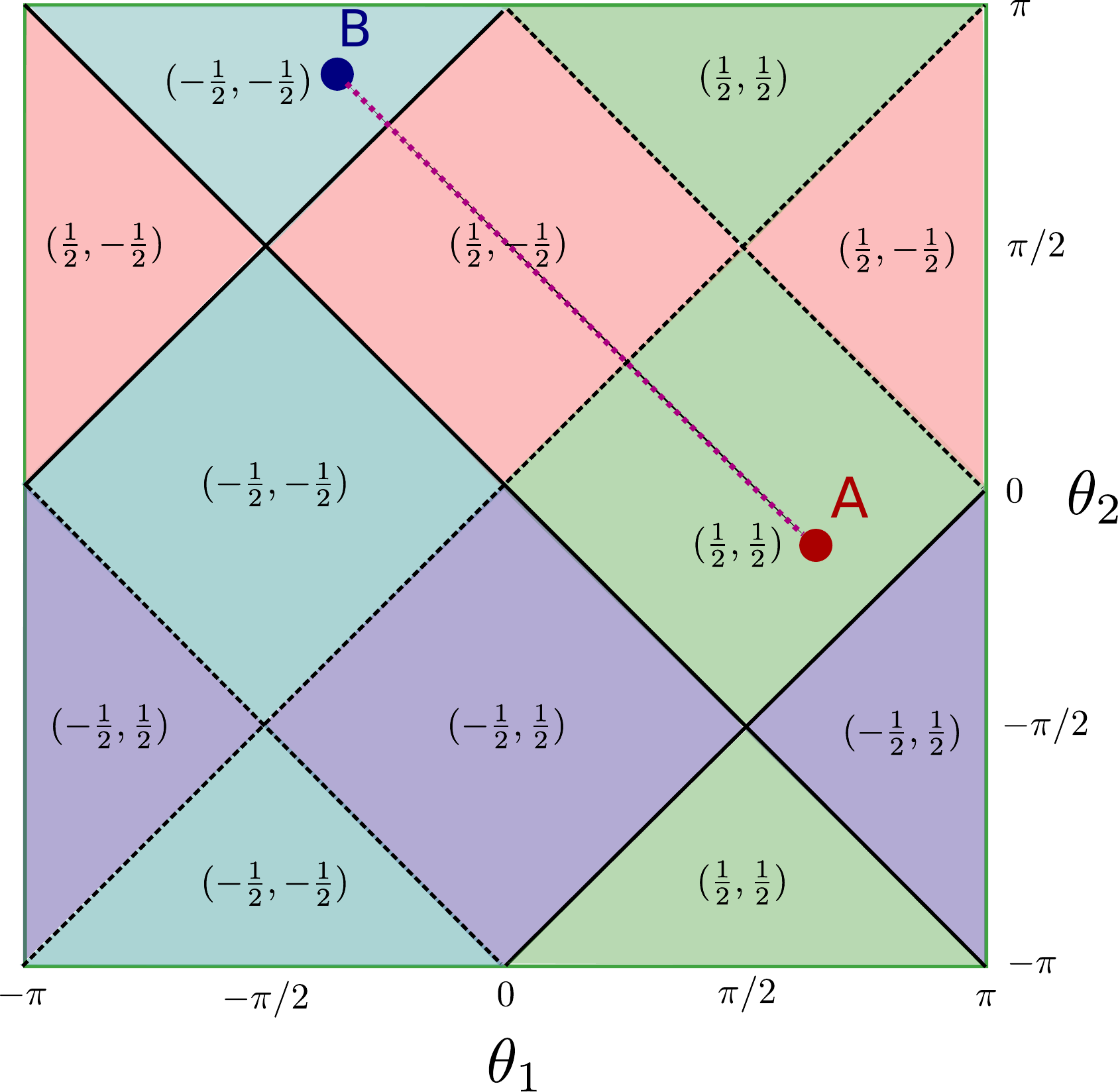}
\caption{(Color online) Split-step walk with binary disorder. The red and blue dots
  represent points $A$ and $B$ in the case $m=1$, while the dashed
  line shows the possible values of $B$, while $m$ goes from $0$ to
  $1$.}
\label{fig:phasemap_binary}
\end{figure}

In the main text we used uniform disorder as an illustration of the
general formalism. In this section we present another simple example
that can be treated analytically and is useful to obtain intuition
regarding the disordered quantum walk. Consider a split-step quantum
walk where the rotation angles $\theta_1$ and $\theta_2$ can take two
different values: $\theta_1^A$, $\theta_2^A$ or $\theta_1^B$,
$\theta_2^B$. At each site, we chose one of these two set of values,
with probabilities $q$ and $1-q$, so that the corresponding
probability measure is
\begin{multline}
\label{eq:binary_def}
\mu(\theta_1,\theta_2) = q\, \delta(\theta_1 - \theta_1^A)
\delta(\theta_2 - \theta_2^A) \\
+ (1 - q)\, \delta(\theta_1 - \theta_1^B)\delta(\theta_2 - \theta_2^B).
\end{multline}
We fix the point $A$ in parameter space as
\begin{align}
\theta_1^A &= 0.625\pi;&
\theta_2^A &= -0.125\pi,
\end{align}
and choose point $B$ to lie on a straight line that is parallel to one
of the phase borders and goes through point $A$,
\begin{align}
\theta_1^B &= \theta_1^A - m\;\pi; &
\theta_2^B &= \theta_2^A + m\;\pi,
\end{align}
where the parameter $m \in [0,1]$ measures the distance between points
$A$ and $B$ in parameter space as shown in
Fig.~\ref{fig:phasemap_binary}. When $m = 0$, the two points coincide
while $m = 1$ corresponds to the case when their distance is maximal.

\begin{figure}[h!]
 \centering
 \includegraphics[width=0.8\columnwidth]{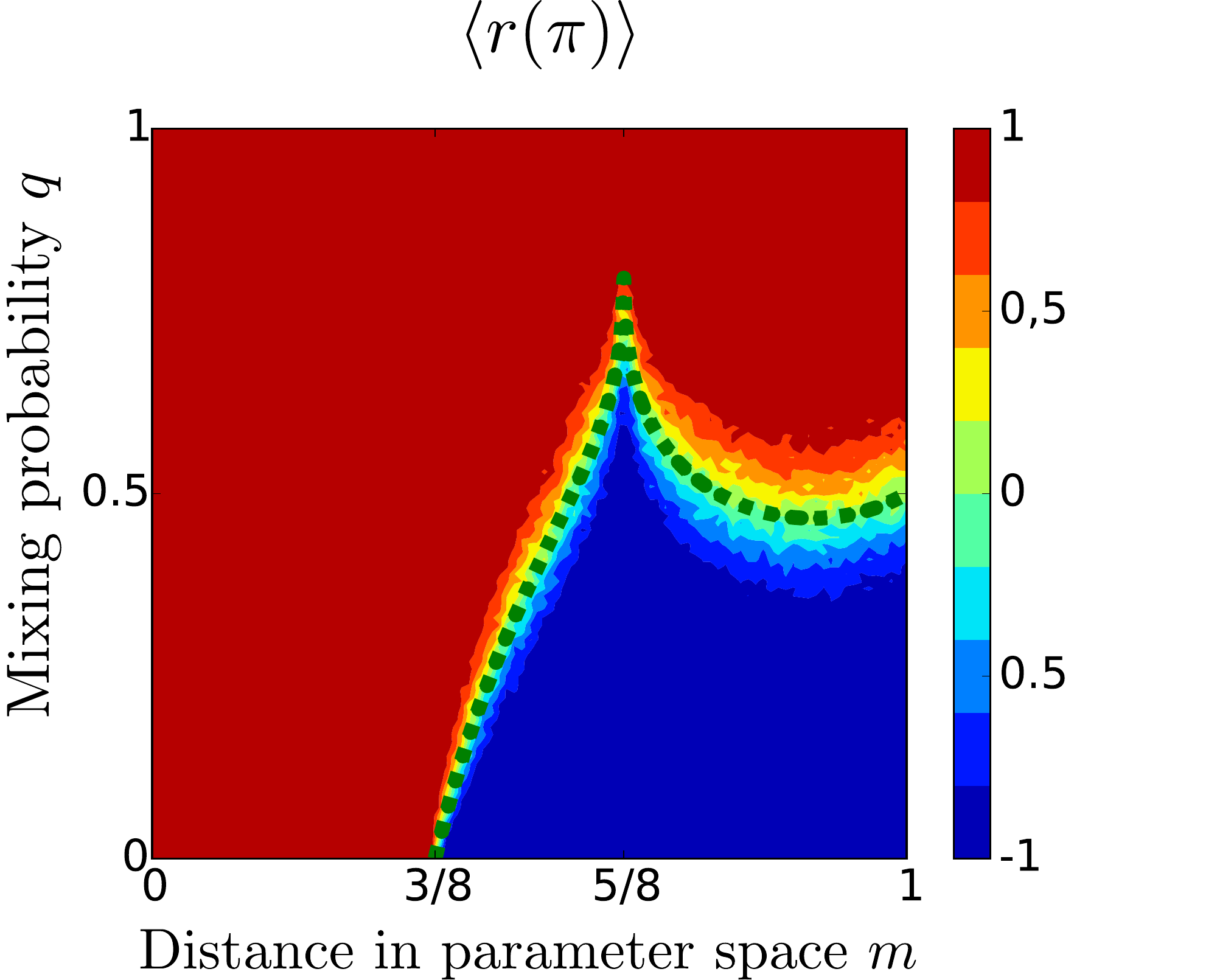}
 \caption{(Color online) Topological phases of the split-step walk with binary
   disorder. Color coding shows the reflection amplitudes at $E=\pi$
   for a system of $40$ sites averaged over $100$ samples. The green
   dashed line shows the critical mixing probability $q_\text{crit}$
   calculated using Eq.~\eqref{eq:qcrit}. The peak at $m = 5/8$
   corresponds the the case when the point $B$ is in the middle of a
   topological phases and $\lambda_\pi$ diverges.}
 \label{fig:qw_binary_phases}
 \end{figure}

At $m = 3/8$, the point $B$ crosses the line where the gap at $E =
\pi$ closes and the invariant $\nu_\pi$ changes. Thus, for values $m >
3/8$ the two limits $q = 0$ and $q = 1$ belong to two different
topological phases. We want to find the exact value $q_\text{crit}$ for
each $m$ where the phase transition occurs. Suppose that we have a
number of $L_A$ sites with parameters $\theta_{1,2}^A$ in our system
and a number of $L_B$ sites from the point $B$. The condition for the
phase transition can then by written as
\begin{equation}
\lambda_E = L_A \lambda_E^A + L_B \lambda_E^B = 0.
\end{equation}
From this we acquire the critical value of the mixing probability $q$, 
\begin{equation}
\label{eq:qcrit}
q_\text{crit} = \frac{1}{1 + \frac{L_B}{L_A}} = 
\frac{1}{1 - \frac{\lambda^A}{\lambda^B}}.
\end{equation}
The critical line $q_\text{crit}(m)$ for $E = \pi$ is shown in
Fig.~\ref{fig:qw_binary_phases} along with the numerically calculated
average reflection amplitudes at quasienergy $\pi$. 

This binary disordered model, while somewhat unphysical, shows the
importance of the Lyapunov exponents defined in
Eqs.~\eqref{eq:tmatrix_parametrization}. They serve as weight factors,
determining how much a given site contributes to the overall
topoogical invariants of the whole system.

\section{Critical exponent for uniform disorder}
\label{sec:critical_exponent}

In this section we will give the critical exponent with which the
localization length $\xi(0)$ or $\xi(\pi)$ diverges when we approach
the topological phase boundary in parameter space. There is also a
different critical exponent when we are at a phase boundary: then the
function $\xi(E)$ diverges as the quasienergy approaches $0$ (or
$\pi$). This was discussed in section \ref{sec:uniform} of the main text.

Let us first look at the clean case. Near the phase boundary $\theta_1
= -\theta_2$ we can expand the function $\lambda_0$ in the small
parameter $\delta = \theta_1 + \theta_2$. From the definition
Eq.~\eqref{eq:Lyapunov_exponents} we get that $\lambda_0 \propto
\delta$ so that $\xi(0) \propto \delta^{-1}$. This means that in the
clean system the critical exponent is $\nu = -1$.

\begin{figure}[h!]
\centering
\includegraphics[width=0.47\columnwidth]{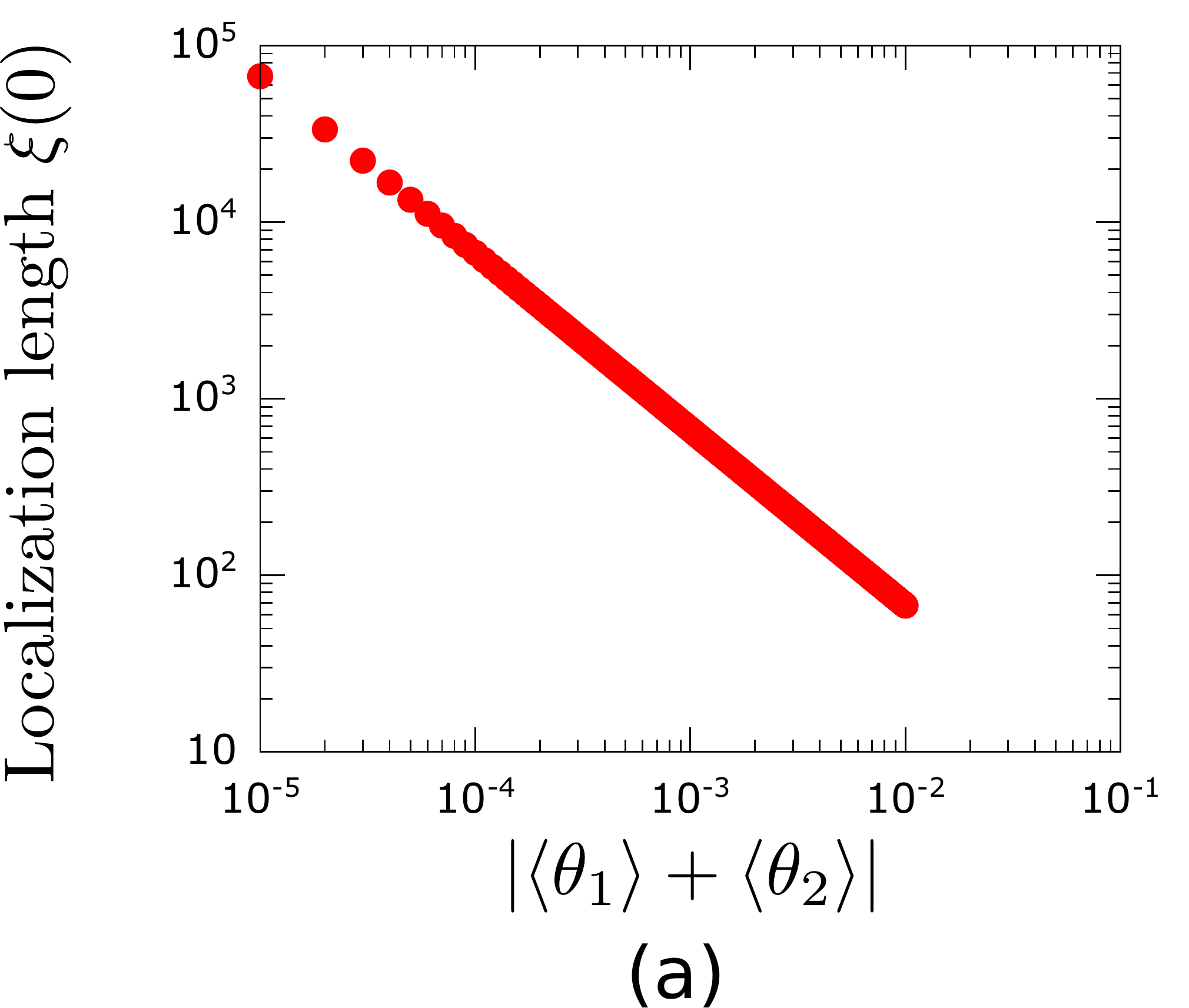}%
\includegraphics[width=0.47\columnwidth]{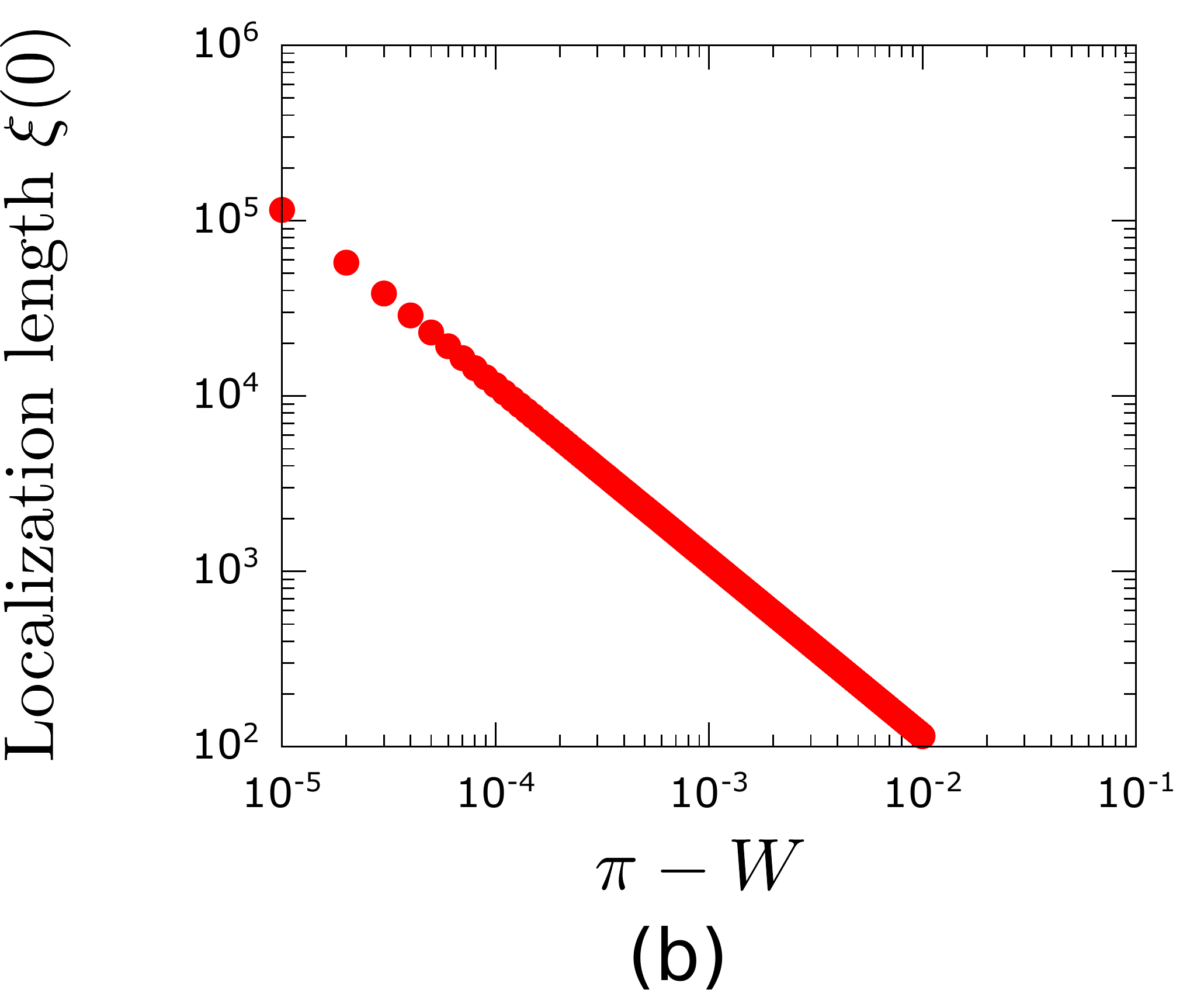}
\caption{Behavior of the localization length $\xi(0)$ near the
  topological phase boundary. (a) Split-step walk with uniform
  disorder, disorder strength set to $W = 0.1\pi$. The average
  rotation angles approach the gap closing line $\theta_1 + \theta_2 =
  0$. (b) Near the strong disorder limit, $W \approx \pi$. In both
  cases, we found a critical exponent $\nu = -1$.}
\label{fig:crit_exponents}
\end{figure}

When uniform disorder is applied, we can approach the phase boundary
in two different ways as shown in Fig.~\ref{fig:qw_uniform_maps}:
either by changing the average values the rotation angles (similarly
to the clean case) or by increasing the disorder strength until we
reach the strong disorder limit as described in the main
text. Numerical evaluation of the integral~\eqref{eq:lambda_avg} in
the vicinity of the phase boundary verifies that the critical exponent
remains $\nu = -1$ in both cases as seen in
Fig.~\ref{fig:crit_exponents}.

\section{Comparison with the real-space winding number method}
\label{sec:realspace_winding}

As we mentioned in the main text, there is an alternative way to
calculate topological invariants for a disordered system, apart from
the scattering matrix approach that we used to study the split-step
walk. This other approach, based on non-commutative geometry tools,
was used to study two dimensional topological insulators at strong
disorder\cite{Prodan_noncommutative}, and, more recently, the
disordered SSH model\cite{mondragon2013topological}. In this
formalism, the winding number of a one dimensional topological
insulator with only two bands can be calculated by the expression
\begin{equation}
\label{eq:realspace_winding}
\nu = -Tr(Q_{-+}[X,Q_{+-}])/L,
\end{equation}
where $X$ is the position operator, $Q_{-+} = \Gamma_- Q\Gamma_+$ and
$Q_{+-} = \Gamma_+ Q\Gamma_-$, with $Q$ being the flat band version of
the Hamiltonian $H$, defined by replacing each eigenvalue of $H$ with
its sign: $Q = \text{sign}(H)$. The operators $\Gamma_+$ and
$\Gamma_-$ are the projectors associated to the chiral symmetry
operator $\Gamma$ by $\Gamma = \Gamma_+ - \Gamma_-$.

Evaluation of the winding number, Eq.~\eqref{eq:realspace_winding},
for a finite size system with periodic boundary conditions requires
non-trivial approximations as described in Prodan's
paper\cite{Prodan_noncommutative}. In this case, the resolution in
quasimomentum space for a system of $L$ sites is given by $\Delta =
2\pi/L$. We approximate the differential $\partial_k A(k)$ for some
function $A(k)$ of the quasimomenta with discrete differences in the
following way:
\begin{equation}
\label{eq:kdiff_discretization}
\partial_k A(k) \rightarrow \delta_k A(k_n) = 
\sum_{m=1}^q c_m[A(k_n + m\Delta) - A(k_n - m\Delta)],
\end{equation}
where $q$ is of order $L/2$.

If we assume that $A(k)$ can be expressed as a Fourier series, then it
is enough to find the coefficients $c_m$ that give a good
approximation for functions of the form $e^{ikx}$. For these, the
formula~\eqref{eq:kdiff_discretization} gives
\begin{equation}
\label{eq:discretization_criteria}
(\partial_k - \delta_k)e^{ikx} = i(x - 2\sum_m c_m \sin(m\Delta x))e^{ikx}.
\end{equation}
We can make the above expression disappear in the first $2q - 1$
orders of $\Delta x$ by choosing $c_m$ to be the solutions of the
equation
\begin{multline}
\boldsymbol{M}\boldsymbol{c}^T = \frac{1}{2\Delta} (1,0,0,\dots,0)^T; \\ 
M_{nm} = m^{2n-1} \qquad n,m = 1,\dots,Q.
\end{multline}
In this case, $(\partial_k - \delta_k)e^{ikx} =
~\mathcal{O}(\Delta^{2q-1})$. By choosing $q=L/2$ we can make the
error of the approximation to be of order
$~\mathcal{O}(\Delta^{L-1})$.

\begin{figure}[h!]
 \centering
 \includegraphics[width=0.45\textwidth]{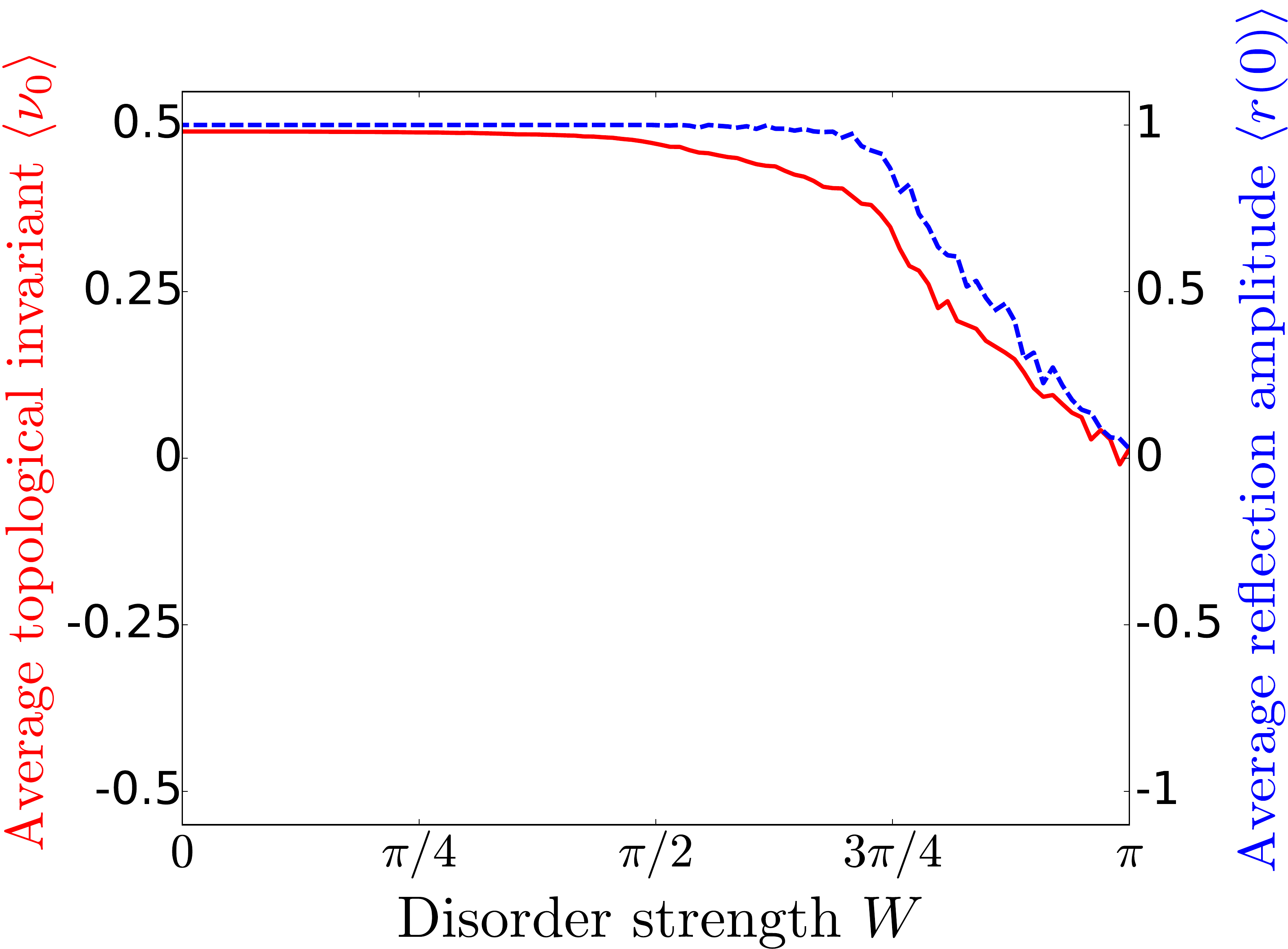}
 \caption{(Color online) For a disordered quantum walk, the bulk
   winding number (solid red) and the average reflection amplitude
   (dashed blue) give similar predictions about the topological phase.
   We used a system of 100 sites and averaged over 1000 disorder
   realizations. For the bulk winding number,
   Eqs.~\eqref{eq:realspace_winding} were used, while the reflection
   amplitude was obtained by iterating the scattering matrix.}
 \label{fig:winding_by_disorder_strength}
 \end{figure}

For the split-step walk, Eq.~\eqref{eq:realspace_winding} can be used
in both chiral symmetric timeframes defined in
Eqs.~\eqref{eq:Up_splitstep_def}
and~\eqref{eq:Upp_splitstep_def}. This gives us two winding numbers
$\nu'$ and $\nu''$. The topological invariants are given as
combinations of these two\cite{asboth_2013}:
\begin{subequations}
\label{eq:invariants_from_winding}
\begin{align}
	\nu_0 &= \frac{\nu' + \nu''}{2}; \\
	\nu_\pi &= \frac{\nu' - \nu''}{2}.
\end{align}
\end{subequations}
These equations, along with Eq.~\eqref{eq:realspace_winding} enable us
to calculate the topological invariants numerically. Below, in
Fig.~\ref{fig:winding_by_disorder_strength} we show the results for
$\nu_0$, along with those performed using the scattering matrices
described in the main text. The two methods yield the same results
qualitatively. However, we note that while the reflection amplitude of
the finite system remains close to the value it has in the
thermodynamic limit even for relatively strong disorder, the
real-space winding number is more easily affected by numerucal
inaccuracies resulting from small system sites. The scattering matrix
method is also more efficient numerically (it scales linearly with the
system size).

\end{document}